\documentclass[journal,10pt]{IEEEtran}

\normalsize

\usepackage{graphicx}
\usepackage{epstopdf}

 \DeclareGraphicsExtensions{.pdf}

\usepackage[cmex10]{amsmath}
\usepackage{bm}
\usepackage{epsfig}
\usepackage{booktabs} 
\usepackage{latexsym}
\usepackage{enumerate}
\usepackage{float} 
\usepackage{subfig}
\usepackage{multirow}
\usepackage{epstopdf}
\usepackage{longtable}
\usepackage{stfloats}
\usepackage{color} 
\usepackage{amssymb}
\graphicspath{{./Figures/}}
\usepackage{color}
\usepackage{tikz}
\usepackage{bbm}
\usepackage{authblk}
\usepackage{verbatim}
\usepackage{mathtools}

\usepackage{balance}

\usepackage{siunitx}
\DeclareSIUnit\bps{bps}
\DeclareSIUnit\Torr{Torr}
\DeclareSIUnit\torr{Torr}
\DeclareSIUnit\sample{Sa}
\DeclareSubrefFormat{parens}{#1(#2)}
\usepackage{cite}

\usepackage{color, soul}
\newcommand{\rev}[1]{\textcolor{black}{{#1}}}

\newtheorem{prop}{Proposition}
\begin{document}

\title{A Framework of Mahalanobis-Distance Metric with Supervised Learning for Clustering Multipath Components in  MIMO Channel Analysis}

\author{Yi~Chen,~\IEEEmembership{Student~Member,~IEEE},  Chong~Han,~\IEEEmembership{Member,~IEEE}, Jia~He, and Guangjian~Wang 
\vspace{0cm}

\thanks{Y. Chen and C. Han are with the Terahertz Wireless Communications (TWC) Laboratory, Shanghai Jiao Tong University, Shanghai, China (email: \{yidoucy, chong.han\}@sjtu.edu.cn).}
\thanks{J. He and G. Wang are with Huawei Technologies Co., Ltd, China (email: \{hejia83, wangguangjian\}@huawei.com).}
}
\markboth{}{}  
\maketitle

\begin{abstract}

As multipath components (MPCs) are experimentally observed to appear in clusters, cluster-based channel models have been focused in the wireless channel study. However, most of the MPC clustering algorithms for MIMO channels with delay and angle information of MPCs are based on the distance metric that quantifies the similarity of two MPCs and determines the preferred cluster shape, greatly impacting MPC clustering quality. In this paper, a general framework of Mahalanobis-distance metric is proposed for MPC clustering in MIMO channel analysis, without user-specified parameters. Remarkably, the popular multipath component distance (MCD) is proved to be a special case of the proposed distance metric framework. Furthermore, two machine learning algorithms, namely, weak-supervised Mahalanobis metric for clustering and supervised large margin nearest neighbor, are introduced to learn the distance metric. To evaluate the effectiveness, a modified channel model is proposed based on the 3GPP spatial channel model to generate clustered MPCs with delay and angular information, since the original 3GPP spatial channel model (SCM) is incapable to evaluate clustering quality. Experiment results show that the proposed distance metric can significantly improve the clustering quality of existing clustering algorithms, while the learning phase requires considerably limited efforts of labeling MPCs.
\boldmath

\end{abstract}

\section{Introduction}
 
In the study of wireless channels, electromagnetic (EM) waves interact with objects in an environment, which thereby propagate through multiple paths including the line-of-sight path, reflected paths, scattering paths, and diffraction paths~\cite{chong2017thz}. These paths are called multipath components (MPC), which are characterized by various physical parameters, e.g., delay, power, angle of arrival (AOA), angle of departure (AOD), phase and etc. Furthermore, MPCs are observed to appear in groups or \textit{clusters} in many channel measurements~\cite{molisch2014propagation}. In a cluster, MPCs generally share similar MPC parameters. 
As a result, the channel characteristics can be divided into \textit{inter-cluster} and \textit{intra-cluster} ones. Separately modeling the inter-cluster and intra-cluster characteristics reduces the channel model parameters and demonstrates superiority in fitting the measurement data. As a result, the concept of cluster constitutes the basis of channel models including SV model~\cite{saleh1987statistical}, Third Generation Partnership Project (3GPP) spatial channel models (SCM)~\cite{3gpp873,3gpp2018study}, Wireless World Initiative New Radio (WINNER) channel models~\cite{WINNER}, Cooperation in the field of Scientific and Technical Research (COST) channel models~\cite{molisch2006cost259,liu2012cost}, Mobile and Wireless Communications Enablers for the Twenty-twenty Information Society (METIS) channel model~\cite{nurmela2015deliverable}, among others. 
\par To this end, MPC clustering algorithms that make use of the delay, power, and angular information to accurately and efficiently gather the MPCs into clusters are the fundamental of channel modeling~\cite{he2018clustering}. In the literature, MPC clustering algorithms generally rely on the distance metric, which quantifies the similarity between two MPCs. In particular, the distance-metric-based MPC clustering algorithms are broadly divided into two categories, namely, partitioning-based algorithms and density-based algorithms.
On one hand, the partitioning-based algorithms determine the clusters by optimizing a criterion related to the distance metric of MPCs. In this category, the KMeans algorithm is classically well-known, which minimizes the sum of squared error of the distance among the MPCs within a cluster~\cite{wagstaff2001constrained}. KPowerMeans further incorporates the power into the KMeans algorithm when determining cluster centroids~\cite{czink2006framework}. Fuzzy-c-means (FCM) algorithm is used in~\cite{schneider2009clustering} and reported to outperform KPowerMeans. The authors in~\cite{huang2019framework} propose a space-transformed fuzzy-c-means (ST-FCM) to address the non-converge problem in~\cite{schneider2009clustering}. Partitioning-based algorithms require the number of clusters as prior information and are sensitive to the initial cluster centroids. On the other hand, density-based clustering algorithms that partition the MPCs according to the density of a local area do not demand the number of clusters and the cluster centroids. For instance, DBSCAN is the most well-known density-based clustering algorithm~\cite{ester1996density}. Kernel-power density-based (KPD) algorithm is proposed by incorporating kernel function based on the cluster characteristics in estimating the density of MPCs~\cite{he2017kernel}. Density-based algorithms do not require the number of clusters as prior information and have the ability to find arbitrary cluster shapes. The minimum number of MPCs in a cluster and the radius that determines the neighborhood of an MPC are the parameters specified by users. 
\par Distinctive from all aforementioned, there are MPC clustering algorithms that are not distance-metric based. These clustering algorithms consider the delay and power of MPCs, and unfortunately, ignore angle information of MPCs which are of significant interest in multiple-input-multiple-output (MIMO) channel models~\cite{rusek2012scaling}. Furthermore, the prior information of cluster features is the premise, including the relation between power and delay, and the distribution of MPC amplitude. For example, the authors in~\cite{shutin2004cluster} use the hidden Markov model to learn the parameters of the MPCs’ distribution and cluster the MPCs in the CIRs. Region competition algorithm along with amplitude distribution of MPCs is used to cluster MPCs by using the Kurtosis measure~\cite{gentile2013using}. The authors in \cite{chuang2007automated} fit a series of exponential curves to the measurements to find clusters. In~\cite{he2016clustering}, a sparsity-based method is proposed to cluster MPCs, regarding that power of the MPCs is exponentially decreasing with delay. Though exploiting the features of clusters is useful to find clusters, one should notice that the prior information of cluster characteristics is a chicken-and-egg challenge, if a good clustering algorithm itself is still not accessible.

\par MPC distance metric determines the preferred shape of clusters, which is critical to the distance-metric-based MPC clustering algorithms. Hence a good distance metric that matches well with the shape of measured clusters can significantly improve the clustering quality. For MPC clustering problems, multipath component distance (MCD)~\cite{steinbauer2002quantify} is proposed to measure the distance between MPCs and reported to outperform the Euclidean distance. In MCD, the delay weighting factor that controls the weight of the delay component is user-specified. Therefore, an open problem is how to find an improved MCD that can automatically adjust to the shape of measured clusters. 

In this paper, we introduce distance metric learning to enhance MPCs clustering quality of the distance-metric-based MPC clustering algorithms. Specifically, we first propose a general framework of the Mahalanobis-distance metric, which can be described by a matrix $\mathbf{A}$, to calculate the similarity of two MPCs. In particular, MCD, which is the most popular distance metric for MPCs clustering in the literature, is proved as a special case of the proposed distance metric. The proposed distance metric shows good compatibility with the existing clustering algorithms and can learn the cluster features from the very limited number of MPCs with labels. In our experiments, the proposed distance metric is found to significantly improve the performance of clustering algorithms, including K-means, KPowerMeans, and DBSCAN. The main contributions of our work are summarized as follows.
\begin{itemize}
    \item We propose a general framework of Mahalanobis-distance metric with distance metric learning for MPC clustering in MIMO channel modeling. Remarkably, we prove that the well-known MCD is a special case of the proposed framework. The proposed distance metric for MPCs presents good compatibility to be incorporated in the existing clustering algorithms without the overhead of modification.
    
    \item To learn the proposed distance metric, we analyze two machine learning methods with different supervision, namely,  weak-supervised \textit{Mahalanobis metric for clustering (MMC)}  and supervised \textit{large margin nearest neighbor (LMNN)}. Furthermore, the two methods require a substantially limited number of labeled samples for learning, i.e. 25 MPCs in one snapshot, which makes them promising for practical implementation.
    
    \item To evaluate the effectiveness of the clustering methods, we find that the 3GPP spatial channel models (SCM) are incapable, which although are used as reference MIMO channel models to generate MPCs with ground-truth labels in the literature. Therefore, we propose a modified model based on the current 3GPP spatial channel model to generate clustered MPCs with delay and angular information. Extensive experiments results illustrate that the proposed general framework of the distance metric can significantly improve the performance of KMeans, KPowerMeans, and DBSCAN algorithms.
    \item Measured channel data obtained from a THz channel measurement campaign from 130~GHz to 143 GHz in a meeting room is used to validate the proposed framework of distance metric. The numerical result shows that the proposed distance metric improves the clustering quality in the sub-THz frequency band by comparing with the MCD.
\end{itemize}
\par The remainder of this paper is organized as follows. Sec.~II describes the MCD-based clustering and the motivation of this work. Sec.~III presents the proposed framework of the Mahalanobis-distance metric for MPC clustering and machine learning methods to learn the proposed metric. In Sec.~IV, a modified MIMO channel model based on the 3GPP SCM is proposed. Based on the proposed MIMO channel model, the performance of the proposed distance metric is analyzed in Sec. V. The proposed distance metric framework is validated by channel measurement data from 130~GHz to 143~GHz in Sec. Vi. Finally, this paper is concluded in Sec. VII.
\section{Multipath Component Distance (MCD)-based Clustering and Remaining Problems}
In this section, we investigate the MPC clustering problem associated with channel measurement and modeling, based on the MCD measure. Moreover, the distance metric in MPC clustering is explained and remaining problems are enlightened.
\subsection{Cluster-based Channel Model}
\par To analyze EM propagation, MPCs are observed to appear in clusters in both delay and angle domains. This observation is justified as an object may cause specular reflected paths and multiple scattered paths. Indeed, the 3GPP SCM, WINNER and COST channel models adopt the cluster as a basic feature of the channel, by separately modeling the inter-cluster and intra-cluster channel statistics. For example, a cluster-based double-directional channel impulse response is described as~\cite{han2016multi}
\begin{equation}
\begin{split}
h(\tau,\phi_T,\theta_T,\phi_R,\theta_R)&=\sum_{n=1}^N\sum_{m=1}^{M_n}\alpha_{n,m}\delta(\tau-\tau_{n,m})
\\&\cdot\delta(\phi_T-\phi_{T,n,m})\delta(\theta_T-\theta_{T,n,m})
\\&\cdot\delta(\phi_R-\phi_{R,n,m})\delta(\theta_R-\theta_{R,n,m})
\end{split},
\end{equation}
where $N$ denotes the number of clusters and $M_n$ represents the number of MPCs in $n^{\text{th}}$ cluster. In addition, $\alpha_{n,m}$ describes the complex amplitude gain of the $m^{\text{th}}$ MPC in the $n^{\text{th}}$ cluster. $\tau_{n,m}$, $\phi_{T,n,m}$, $\theta_{T,n,m}$, $\phi_{R,n,m}$ and $\theta_{R,n,m}$ stand for the time-of-arrival (TOA), azimuth angle of departure (AAOD), zenith angle of departure (ZAOD), azimuth angle of arrival (AAOA) and zenith angle of arrival (ZAOA) of the $m^{\text{th}}$ MPC in the $n^{\text{th}}$ cluster, respectively. In this paper, the parameter set of the $l^\text{th}$ MPC is denoted by
\begin{equation}
    \mathbf{\omega}_l=\{\alpha_l,\tau_l,\phi_{T,l},\theta_{T,l},\phi_{R,l},\theta_{R,l}\}.
\end{equation} 
\par Moreover, the set of all the MPCs in one snapshot is denoted by $\Omega=\{\mathbf{\omega}_l|l=1,2,\dots,L\}$ where $L$ is the total number of MPCs in a snapshot.
\subsection{MPCs Extraction from Channel Measurements}
\par In channel measurement, time-domain and frequency-domain sounding methodologies both measure the power and delay of MPCs, with the following difference. The time-domain channel sounding directly measures the channel impulse response (CIR), while the frequency-domain channel sounding records channel transfer function (CTF) with limited bandwidth. CIR can be then calculated by performing inverse Fourier transform of CTF, which however causes that the MPCs exceeding the maximum detectable delay are wrapped in the delay domain. 

\par Besides the temporal domain, two methods are presented in extracting MPCs in the angular domain. First, MIMO techniques with multiple antennas provide the angle resolution of MPCs. However, as RF chains are expensive, a virtual antenna array enabled by antenna shifting is a widely-used alternative method in the practice of MIMO channel measurements. Unfortunately, the virtual antenna array ignores mutual coupling effects among antennas~\cite{zhang2020automatic}. For MIMO channel post-processing, SAGE algorithms can be used to extract MPCs with delay and angle information from measured MIMO channels. 
In addition to MIMO techniques, rotating directional antennas to scan the angle domain is another method to acquire angular information of MPCs~\cite{yu2020wideband}. For the channel measurement with directional antenna scanning, it is still an open problem to extract accurate MPCs from the measured directional channels. Although a modified SAGE algorithm is proposed to estimate MPCs in the channel measurement with rotation of directional antenna~\cite{yin2016performance}, the proposed algorithm is sensitive to phase error due to the swing cable during antenna rotation and requires an accurate three-dimension radiation pattern of the directional antenna at each rotated angle~\cite{60ghzWCX}. In practical post-processing, an over-simplified MPC extraction approach is widely adopted, which simply regards the rotated angles of Tx and Rx as the angles of departure and arrival, respectively~\cite{60ghzWCX}.

\subsection{Multipath Component Distance (MCD) Metric}
Currently, most of the clustering algorithms for MPCs with angle information are based on the distance metric of two MPCs, which in turn has a great impact on the clustering algorithm performance. Then, the preferred cluster shape is decided by the distance metric of MPCs. The widely-adopted distance metrics for data clustering include Manhattan distance, Euclidean distance, and Chebyshev distance. Correspondingly, the preferred clusters of these distance metrics are rhombus shape, ring shape, and square shape, respectively~\cite{he2018clustering}. To improve the clustering quality of Euclidean distance, MCD is further proposed~\cite{steinbauer2002quantify}, for which the MCD between $\mathbf{\omega}_i$ and $\mathbf{\omega}_j$ is calculated as
\begin{equation}
    MCD_{i,j}=\sqrt{ MCD_{\tau,i,j}^2+MCD_{T,i,j}^2+MCD_{R,i,j}^2}
    \label{eq:MCD},
\end{equation}
where $MCD_{\tau,i,j}$ stands for the MCD for TOA, as
\begin{equation}
    MCD_{\tau,i,j}=\xi |\tau_i-\tau_j|,
\end{equation}
where $\xi=\zeta\gamma$ is a factor containing both the delay weighting factor that determines the importance of the TOA, $\zeta$, and the scaling factor that normalizes the TOA, $\gamma$. \rev{One representation of $\gamma$ is $\tau_{std}/\Delta\tau^2_{max}$, where $\Delta\tau_{max}=\max\{
\tau_i-\tau_j\}$ computes the maximum difference between two MPCs and $\tau_{std}$ is the tandard deviation of TOA~\cite{czink2006framework}.}
\par Moreover in \eqref{eq:MCD}, the second and third terms $MCD_{T/R,i,j}$ state the MCD for angle of departure or arrival as
\begin{equation}
\begin{split}
    &MCD_{T/R,i,j}=
    \\&
    \frac{1}{2}\begin{bmatrix}
         \sin\theta_{T/R,i}\cos\phi_{T/R,i} -\sin\theta_{T/R,j}\cos\phi_{T/R,j}  \\
         \sin\theta_{T/R,i}\sin\phi_{T/R,i} -\sin\theta_{T/R,j}\sin\phi_{T/R,j} \\
         \cos\theta_{T/R,i}-\cos\theta_{T/R,j}
    \end{bmatrix}
\end{split}.
\end{equation}
\par MCD noticeably outperforms the Euclidean distance for clustering~\cite{steinbauer2002quantify}, since MCD transforms the angle components of MPCs to the spherical space. This transform avoids the problem caused by Euclidean distance that two MPCs each with azimuth angle of $1^\circ$ and $359^\circ$ are separated apart. 

\subsection{Remaining Problems}
\par Since MCD was proposed in 2006, it has been widely accepted as a distance metric in MPC clustering research. MCD shows that a good distance metric is capable of improving the performance of existing clustering algorithms. 
Although MCD enhances the clustering quality compared to the Euclidean distance, the weight of each component in calculating the distance metric is still a critical issue. For example, increasing the weight of TOA in calculating MCD in~\eqref{eq:MCD}, i.e., the delay scaling factor, the MPCs in a cluster are preferred to be more compact in the TOA domain. Nevertheless, the determination of delay scaling factor is not specific and often manually selected by experience.

Therefore, how to choose an optimal weight of each component in the distance metric, e,g, the delay scaling factor in MCD, to adapt to the cluster shape is still an open problem. One step further, how to find a better distance metric of MPCs is yet to be explored~\cite{he2018clustering}. In light of these, we make our attempt to investigate Mahalanobis-distance metric learning and propose a general framework of distance metric of MPCs as a solution to these two problems. 


%

\section{Distance Metric Learning for MPCs Clustering}
In this section, we present a general framework of Mahalanobis-distance metric for MPCs and prove that MCD is a special case of this framework. Then, we introduce a weak-supervised learning approach and a supervised learning algorithm to learn the proposed distance metric.
\subsection{A General Framework of Mahalanobis-Distance Metric for MPCs}
\par First, we place all the MPCs in a space $\mathbb{R}^7$, by which each MPC is represented by a vector, $\mathbf{x}_l$, as
\begin{equation}
    \mathbf{x}_l=
    \begin{bmatrix}
    \tau_l\\
    \sin\theta_{T,l}\cos\phi_{T,l}\\
    \sin\theta_{T,l}\sin\phi_{T,l}\\
    \cos\theta_{T,l}\\
    \sin\theta_{R,l}\cos\phi_{R,l}\\
    \sin\theta_{R,l}\sin\phi_{R,l}\\
    \cos\theta_{R,l}
    \end{bmatrix}
    \label{eq:xl}.
\end{equation}
    \par Then, we define the Mahalanobis-distance metric between two MPCs, i.e. $\mathbf{x}_i$ and $\mathbf{x}_j$, in the following form~\cite{xing2002distance},
    \begin{equation}
    \begin{split}
        dist(\mathbf{x}_i,\mathbf{x}_j)&=||\mathbf{x}_i-\mathbf{x}_j||_{\mathbf{A}}
        \\&=\sqrt{(\mathbf{x}_i-\mathbf{x}_j)^T \mathbf{A}(\mathbf{x}_i-\mathbf{x}_j) },
        \end{split}
        \label{eq:distance-metric}
    \end{equation}
    \rev{where $||\cdot||_{\mathbf{A}}$ denotes the $\mathbf{A}$ norm, and $\mathbf{A}$ is a real semi-definite matrix satisfying $\mathbf{A}\succeq 0$, which guarantees that the metric satisfies non-negativity and triangle inequality.} To interpret, matrix $\mathbf{A}$ parameterizes a family of Mahalanobis distance over the space of MPCs.
    
    \begin{prop} The MCD in~\eqref{eq:MCD} is a special case of the Mahalanobis-distance metric given in~\eqref{eq:distance-metric}, when $\mathbf{A}$ is a diagonal matrix given as
    \begin{equation}
        \mathbf{A}=\begin{bmatrix}
        \xi^2&0&0&0&0&0&0\\
        0&\frac{1}{4}&0&0&0&0&0\\
        0&0&\frac{1}{4}&0&0&0&0\\
        0&0&0&\frac{1}{4}&0&0&0\\
        0&0&0&0&\frac{1}{4}&0&0\\
        0&0&0&0&0&\frac{1}{4}&0\\
        0&0&0&0&0&0&\frac{1}{4}\\
        \end{bmatrix}
        \label{eq:A}.
    \end{equation}
    \label{prop1}
    \end{prop}
    \par The proof of proposition~\ref{prop1} can be derived by substituting~\eqref{eq:A} and~\eqref{eq:xl} into~\eqref{eq:distance-metric}. Proposition~\ref{prop1} suggests that the proposed Mahalanobis-distance is a more general form of distance metric for MPCs than MCD. \textcolor{black}{Also, the computational complexity of the proposed distance metric is the same as the MCD.}
        \begin{prop} The Mahalanobis-distance metric in~\eqref{eq:distance-metric} is equivalent to the Euclidean distance metric when MPCs are transformed via a transform matrix $\mathbf{A}^{\frac{1}{2}}$, denoted as, $\hat{\mathbf{x}}_l=\mathbf{A}^{\frac{1}{2}}\mathbf{x}_l$.
    \label{prop2}
    \end{prop}
    \par Let $M(\cdot)$ be a mapping function for the MPC set and defined as
    \begin{equation}
        M:\mathbf{x}_l\in \mathbb{R}^7 -> \hat{\mathbf{x}}_l\in \mathbb{R}^7 ,
    \end{equation}
    and
    
    \begin{equation}
        \hat{\mathbf{x}}_l=M(\mathbf{x}_l)=\mathbf{A}^{\frac{1}{2}}\mathbf{x}_l.
        \label{eq:transform}
    \end{equation}
  \par Then, we have the following relation between the Euclidean distance of the transformed MPCs and the Mahalanobis-distance metric of the original MPCs,
  \begin{equation}
          ||\mathbf{x}_i-\mathbf{x}_j||_{\mathbf{A}}=||\hat{\mathbf{x}}_i-\hat{\mathbf{x}}_j||_{2}
  \end{equation}
     \par From the view of this coordinate transformation, the proposed distance metric is well compatible with all the MPC clustering algorithms in the literature, since the adoption of the proposed distance metric is equivalent to implementing the existing clustering algorithms on the transformed MPCs $\hat{\mathbf{x}}_l$. This indicates that the proposed distance metric can be easily incorporated to the existing MPC clustering algorithms without incurring any modification overhead.
     \par To further clarify, the coordinate transformation of MPCs in~\eqref{eq:transform} is linear. The non-linear transformation can be achieved by forwarding the original MPCs through a non-linear basis function $\varphi(\cdot)$ and replacing~\eqref{eq:distance-metric} by $\sqrt{(\varphi(\mathbf{x}_i)-\varphi(\mathbf{x}_j))^T \mathbf{A}(\varphi(\mathbf{x}_i)-\varphi(\mathbf{x}_j)) }$. The basis function $\varphi(\cdot)$ is similar to the idea of kernel function in~\cite{he2017kernel} for improving DBSCAN. The basis function $\varphi(\cdot)$  is manually chosen and highly depends on the cluster parameter statistics of the well-clustered MPCs obtained from the channel measurements, which requires a massive number of ground-truth MPC clustering results. Therefore, we do not consider this non-linear basis function in this work. Furthermore, the amplitude gain of each MPC is not considered in~\eqref{eq:xl} though it can be incorporated like TOA. The reason is that, although the amplitude gain of MPC is involved in the clustering algorithm in the literature, it functions as a weight in calculating the centroids of clusters in KPowerMeans or the kernel function in KPD, which should not be considered as a component of the distance metric.

\subsection{Weak-supervised Learning Algorithm}
To this end, learning the distance metric is equivalent to learn the semi-definite matrix $\mathbf{A}$. We propose a \textit{Mahalanobis metric for clustering (MMC)} method to learn the matrix $\mathbf{A}$~\cite{xing2002distance} by minimizing the loss function, $\mathbf{L}_{\text{MMC}}(\mathbf{A})$, which represents the squared distance between the pairs of MPCs in the same cluster, given by
\begin{equation}
    \mathbf{L}_{\text{MMC}}(\mathbf{A})=\min_{\mathbf{A}}\sum_{(\mathbf{x}_i,\mathbf{x}_j)\in S}||\mathbf{x}_i-\mathbf{x}_j||^2_{\mathbf{A}},
\end{equation} 
where $S$ is a set defined as $(\mathbf{x}_i,\mathbf{x}_j)\in S$, if $\mathbf{x}_i$ and $\mathbf{x}_j$ are in the same cluster. In the learning phase, the required information of the MPCs is whether the two MPCs belong to the same cluster, which is not sensitive to the exact number of clusters. Therefore, we state that MMC is categorized as a weak-supervised learning approach.
\par As $\mathbf{A}=0$ is a trivial solution by grouping all the MPCs into one cluster, an additional constraint $\sum_{(\mathbf{x}_i,\mathbf{x}_j)\in D}||\mathbf{x}_i-\mathbf{x}_j||_{\mathbf{A}}\ge 1$ is introduced, where $D$ has the opposite definition as $S$ which denotes that two MPCs are not in the same cluster. As a result, the optimization problem is formalized as
\begin{subequations}
\begin{align}
        \min_{{\mathbf{A}}}\ &\quad \mathbf{L}_{\text{MMC}}(\mathbf{A}) \\
        \text{s.t.}&\sum_{(\mathbf{x}_i,\mathbf{x}_j)\in D}||\mathbf{x}_i-\mathbf{x}_j||_{\mathbf{A}}-1\ge 0,\\
        &\quad\ {\mathbf{A}}\succeq0
\end{align}
\end{subequations}

\par
\begin{figure}
\centering
\includegraphics[width=0.48\textwidth]{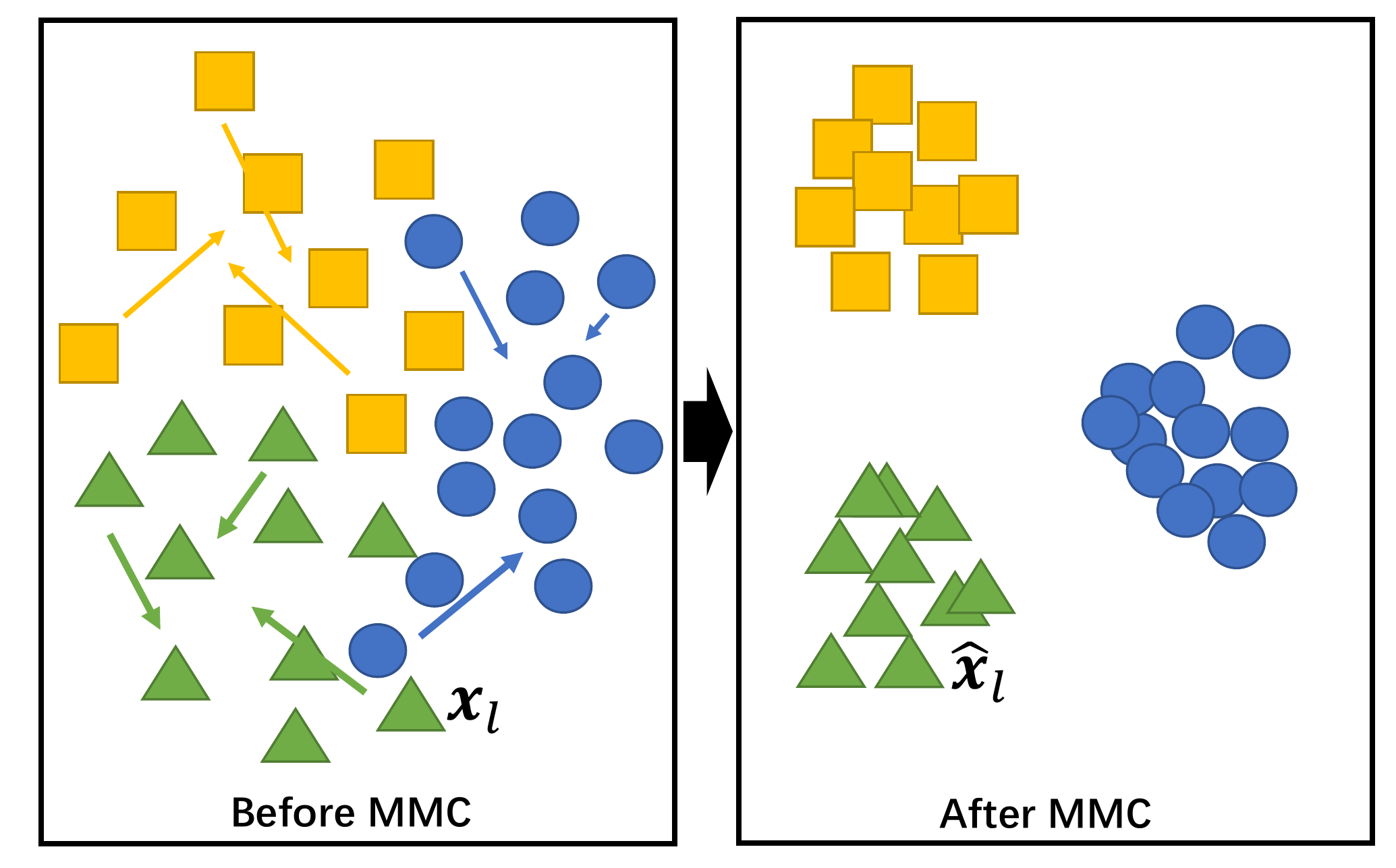}
\caption{Illustration of MMC for clustering. Different shapes and colors represent different clusters. }
\label{fig:mmc}
\end{figure}

 The purpose of this MMC criterion is to make the MPCs of a cluster more compact after a coordinate transformation, which can benefit the clustering quality. An illustration of the MMC learning for clustering is shown in Fig.~\ref{fig:mmc}, where $\mathbf{x}_l$ is the original MPC while $\hat{\mathbf{x}}_i$ is the transformed MPC via $\mathbf{A}$ learned by MMC.

Based on Proposition 1, if we learn the diagonal of $A$ which relates our proposed Mahalanobis-distance metric to MCD, the optimization problem is equivalent to minimize a loss function, $\mathbf{L}_{\text{MMC}}^{\text{diag}}(\mathbf{A})$, as
\begin{equation}
\begin{split}
    \mathbf{L}_{\text{MMC}}^{\text{diag}}(\mathbf{A})=&\sum_{(\mathbf{x}_i,\mathbf{x}_j)\in S}||\mathbf{x}_i-\mathbf{x}_j||^2_{\mathbf{A}}
    \\&-\log \left(\sum_{(\mathbf{x}_i,\mathbf{x}_j)\in D}||\mathbf{x}_i-\mathbf{x}_j||_{\mathbf{A}}\right),
    \end{split}
\end{equation}
which can be efficiently solved by the Newton-Raphson method. We note that the learned diagonal $\mathbf{A}$ can be regarded as to learn the weights of each components in MCD, including not only the delay weighting factor but also the weighting factors for AOA and AOD which are fixed to 0.5 in MCD. 
\subsection{Supervised Learning Algorithm}
MMC builds on the assumption by attempting to minimize distances between all pairs of similarly labeled MPCs, which is reasonable for unimodal MPC clusters. However, real-world MPC clusters cannot be simply modeled as unimodal distributions. Therefore, a more realistic yet accurate objective for distance metric learning is needed to avoid these parametric assumptions about the distribution of MPC clusters.
For this purpose, we analyze a \textit{large margin nearest neighbor (LMNN)} method for clustering, which is originally developed for classification tasks~\cite{weinberger2006distance,weinberger2009distance}. In particular, LMNN is based on the following two intuitions. 
\begin{itemize}
    \item Each training input should share the same label as its $k$ nearest neighbors. 
    \item Training inputs with different labels should be sparsely separated. 
\end{itemize} 
\par In order to learn a transform that the MPCs satisfy these properties, the design of the loss function needs to balance the following two competing terms. On one hand, large distances between nearby inputs with the same labels should be penalized. On the other hand, small distances between inputs with different labels should be penalized meanwhile.
\par In light of these, we first identify that the \textit{target neighbors} of each $\mathbf{x}_i$ are the $K$ nearest neighbors sharing the same label with $\mathbf{x}_i$. As the prior knowledge of the distance metric is currently absent, MCD can be utilized to determine the $K$ nearest neighbors. The notation $j\rightsquigarrow i$ is used to denote that $\mathbf{x}_j$ is a target neighbor of $\mathbf{x}_i$. Then, \textit{impostors} of an arbitrary MPC, $\textbf{x}_i$, are defined,
\begin{equation}
\begin{split}
    \Omega^{\text{impostors}}_{i}:\{\mathbf{x}_l| \  &||\mathbf{x}_l-\mathbf{x}_i||^2_{\mathbf{A}}\le ||\mathbf{x}_j-\mathbf{x}_i||^2_{\mathbf{A}}+1,\\ 
    &j\rightsquigarrow i,y_l\neq y_i\}
    \end{split}
\end{equation}
where $\mathbf{x}_j$ is an any target neighbor of $\mathbf{x}_i$ and $y_l$ is the label of $l^{\text{th}}$ MPC. As a result, an impostor of the MPC, $\textbf{x}_i$, is a MPC with different label that invades a margin defined by a target neighbor of the MPC.
\begin{figure}
\centering
\includegraphics[width=0.48\textwidth]{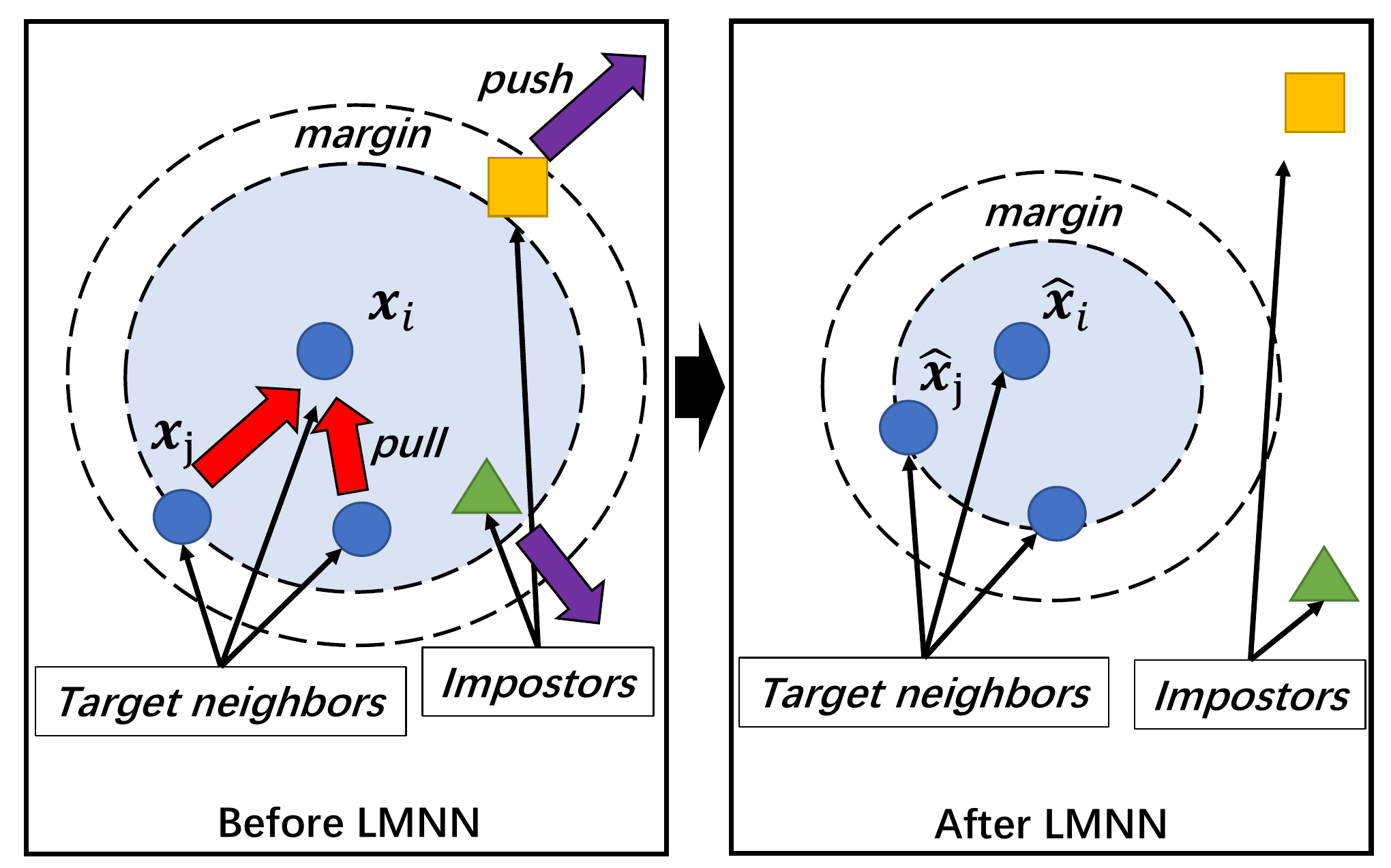}
\caption{Illustration of LMNN for clustering. Different shapes and colors denotes different clusters.}
\label{fig:lmnn}
\end{figure}
\par The key idea of LMNN is to \textit{pull} the target neighbors close to each other, while \textit{pushing} the imposters with different labels away, as shown in Fig.~\ref{fig:lmnn}. In this example, $\mathbf{x}_i$ and $\mathbf{x}_j$ are the original MPCs, while $\hat{\mathbf{x}}_i$ and $\hat{\mathbf{x}}_j$ are the transformed MPCs via $\mathbf{A}$. Learned by LMNN, $\hat{\mathbf{x}}_i$ and $\hat{\mathbf{x}}_j$ are pulled closer, while the imposters (e.g., yellow rectangle and green triangle) are pushed away.
As a result, the loss function, $\mathbf{L}_{\text{LMNN}}(\mathbf{A})$, is a combination of two terms, stated as
\begin{equation}
    \mathbf{L}_{\text{LMNN}}(\mathbf{A})=(1-\mu)\epsilon_{\text{pull}}(\mathbf{A})+\mu\epsilon_{\text{push}}(\mathbf{A}),
    \label{eq:loss_LMNN}
\end{equation}
where $\mu$ is a tunable parameter and set to be 0.5 in practice.
In \eqref{eq:loss_LMNN}, $\epsilon_{\text{pull}}(\mathbf{A})$ and  $\epsilon_{\text{push}}(\mathbf{A})$ denote the single loss function for the actions of pull and push, respectively.
On one hand, $\epsilon_{\text{pull}}(\mathbf{A})$ penalizes the large distance between the MPCs in the same cluster, which is the sum of the squared distances between an MPC and its the target neighbors, given by
\begin{equation}
    \epsilon_{\text{pull}}(\mathbf{A})=\sum_{j\rightsquigarrow i}||\mathbf{x}_i-\mathbf{x}_j||_{\mathbf{A}}^2.
\end{equation}
On the hand, $\epsilon_{\text{push}}(\mathbf{A})$ penalizes the small distance between an MPC and the imposters, as
\begin{equation}
\begin{split}
    \epsilon_{\text{push}}(\mathbf{A})=&\sum_{i,j\rightsquigarrow i}\sum_l\{(1-y_{i,l})\\
    &\cdot\left[1+||\mathbf{x}_j-\mathbf{x}_i||^2_{\mathbf{A}}-||\mathbf{x}_l-\mathbf{x}_i||^2_{\mathbf{A}}\right]_+ \}
    ,\end{split}
\end{equation}
where $y_{i,l}=1$ if $y_i=y_j$, and $y_{i,l}=0$ otherwise. 
\par To minimize the loss function, $\mathbf{L}_{\text{LMNN}}(\mathbf{A})$, a semi-definite program (SDP) is formulated by introducing slack variables $\chi_{i,j,l}$ for all triplets of target neighbors ($j\rightsquigarrow i$) and imposters $\mathbf{x}_l$, as
\color{black}
\begin{subequations}
\begin{align}
\min_{\mathbf{A}} \ &(1-\mu)\sum_{j\rightsquigarrow i}||\mathbf{x}_j-\mathbf{x}_i||_{\mathbf{A}}^2+\mu\sum_{i,j\rightsquigarrow i,l}(1-y_{i,l})\chi_{i,j,l}\\
\mathrm{s.t.} \ & ||\mathbf{x}_l-\mathbf{x}_i||^2_{\mathbf{A}}-||\mathbf{x}_j-\mathbf{x}_i||^2_{\mathbf{A}}-1+\chi_{i,j,l}\ge 0,\\
&\ \chi_{i,j,l}\ge 0,\\
&\ \mathbf{A}\succeq 0.
\end{align}
\end{subequations}
\color{black}
\par The SDP can be solved by a special-purpose solver developed in~\cite{weinberger2009distance}. 

\textbf{Remarks:} To solve the proposed Mahalanobis-distance metric in~\eqref{eq:distance-metric}, two machine learning methods with different supervision are explained in this work, including the weak-supervised MMC and supervised LMNN learning algorithms. 
To differentiate, we use MMC to solve the diagonal matrix $\mathbf{A}$ in Sec.~III-B, while we use LMNN to learn the full matrix $\mathbf{A}$ in Sec.~III-C. The design of the loss function of MMC is relatively simpler than the LMNN and the performance of learned diagonal $\mathbf{A}$ is theoretically poorer than full $\mathbf{A}$. The reason to analyze the diagonal matrix $\mathbf{A}$ is that learning diagonal $\mathbf{A}$ is equivalent to learn the delay weight factor of MCD as well as the weight factors for the components related to AOD and AOD that are fixed values in MCD.

\section{Cluster-based MIMO Channel Models for Evaluating Clustering Quality}
In this section, we first analyze the cluster features of 3GPP SCMs and shows that they are not appropriate as reference MIMO channel models in evaluating clustering quality. Then, we propose some modifications to the 3GPP SCM model to better evaluate clustering quality.
\subsection{3GPP SCM below 100 GHz}
\par When evaluating the performance of clustering algorithms, the importance of datasets can not be underestimated. In the literature, 3GPP SCMs are used as reference MIMO channel models to generate clustered MPCs with power, delay, and angle information for evaluating the clustering quality of MPC clustering algorithms~\cite{czink2006framework,he2017kernel,huang2019framework}. However, 3GPP TR 36.873 channel model~\cite{3gpp873} and 3GPP TR 38.901 channel model~\cite{3gpp2018study}, which are proposed for frequencies from below 100 GHz covering the mmWave band, generate MPCs have the following two features:
\begin{itemize}
    \item The power is equally distributed over the MPCs in a cluster, while the power values of clusters are distinctive. 
    \item Among all the 20 clusters, only 2 of them with the strongest power have the cluster delay spread, i.e., the delays of the intra-cluster MPCs are not identical. Among different clusters, the cluster delays are distinctive. Note that in the early version of the 3GPP channel model~\cite{3gpp996}, none of the clusters has cluster delay spread due to the narrow bandwidth.
\end{itemize}

\par These two cluster features are reasonable for link-level and system-level capacity analysis of wireless communication systems. However, they result in the fact that the generated MPCs can be well clustered according to the power and delays. For example, we can simply cluster the MPCs with the same received power, which is equivalent to use power as the distance metric of MPCs, to get perfect clustering quality. As power is not directly used in distance metrics in the literature, one can use the delay as the distance metric of MPCs and can have good clustering quality, although MPCs in the 2 clusters with the strongest power may be incorrectly clustered. This is equivalent to set the delay weighting factor in MCD as an infinitely large value. Therefore, if one implements the 3GPP SCM model to generate MPCs and takes MCD as the distance metric of MPCs, the clustering algorithm performance can be greatly improved by increasing the delay weighting factor in MCD. Nevertheless, a doubt arises whether the improved clustering quality results from the greater delay weighting factor or the design of new clustering algorithms.
\begin{figure*}[htbp]
\subfloat[Original MPCs, $\mathbf{x}_l$.]{
\includegraphics[width=0.23\textwidth]{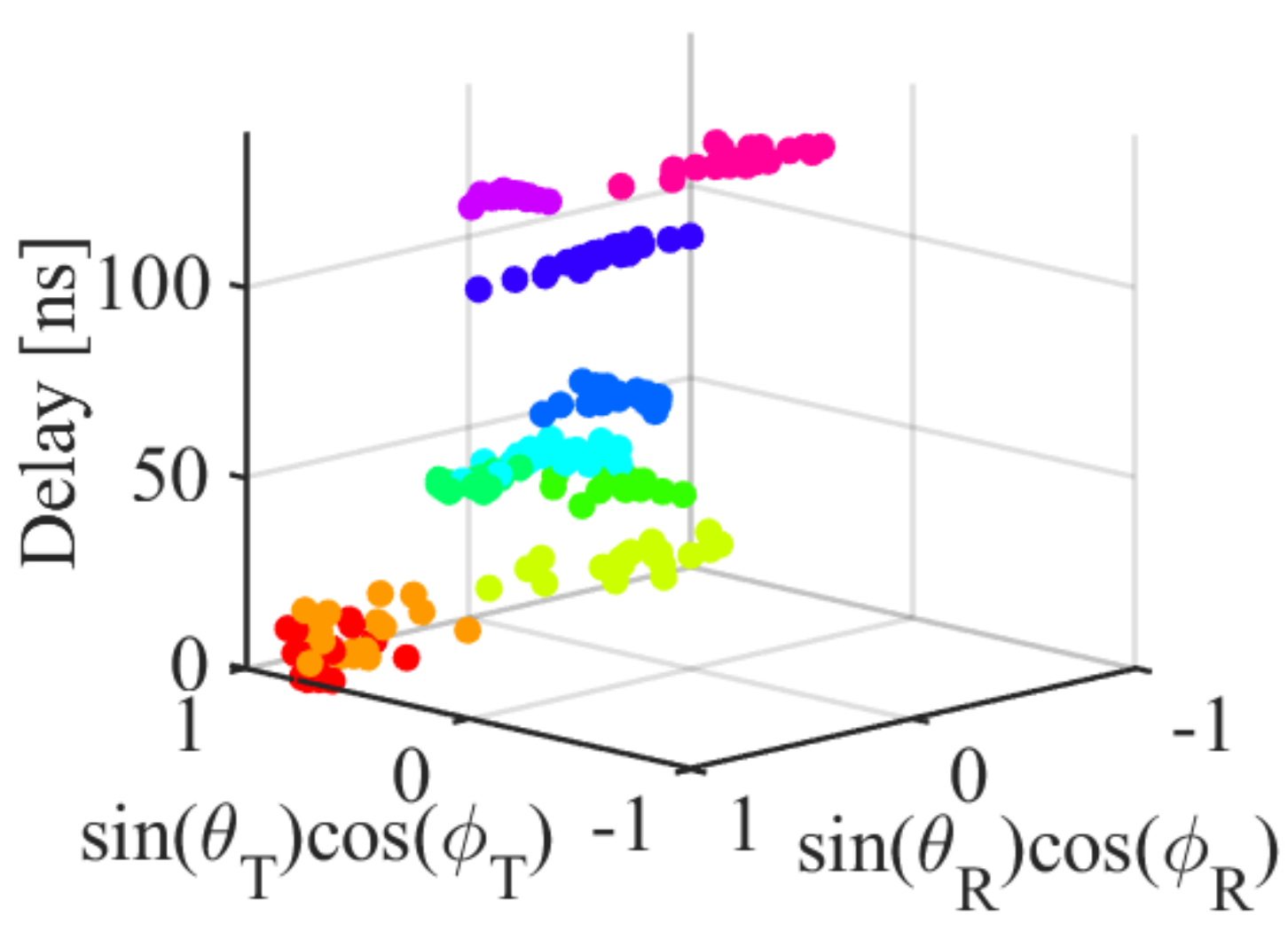}
\label{fig:old_model_1a}
}
\subfloat[Original MPCs, $\mathbf{x}_l$.]{
\includegraphics[width=0.23\textwidth]{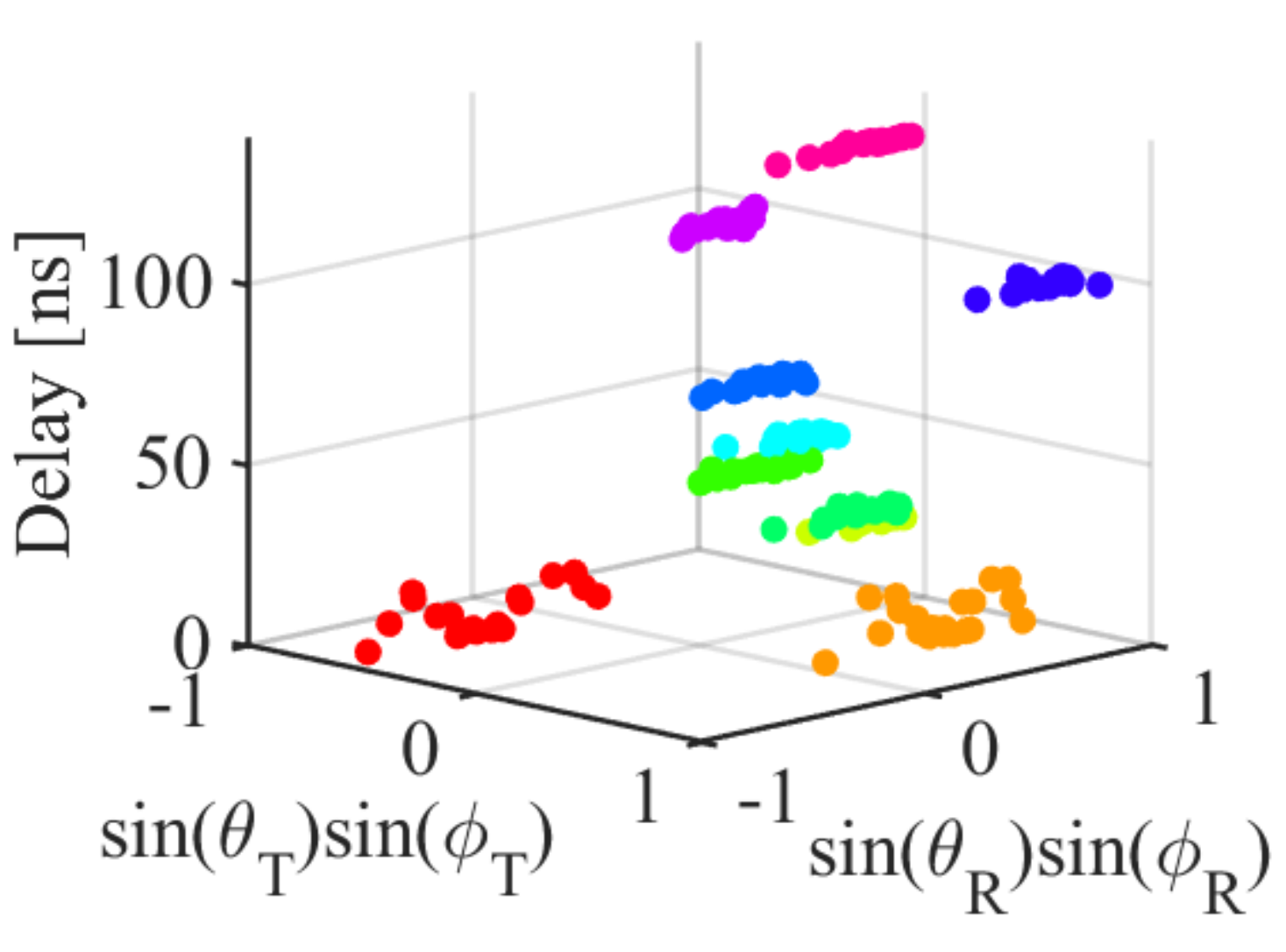}
\label{fig:old_model_1b} 
}
\subfloat[Transformed MPCs, $\hat{\mathbf{x}}_l$.]{
\includegraphics[width=0.23\textwidth]{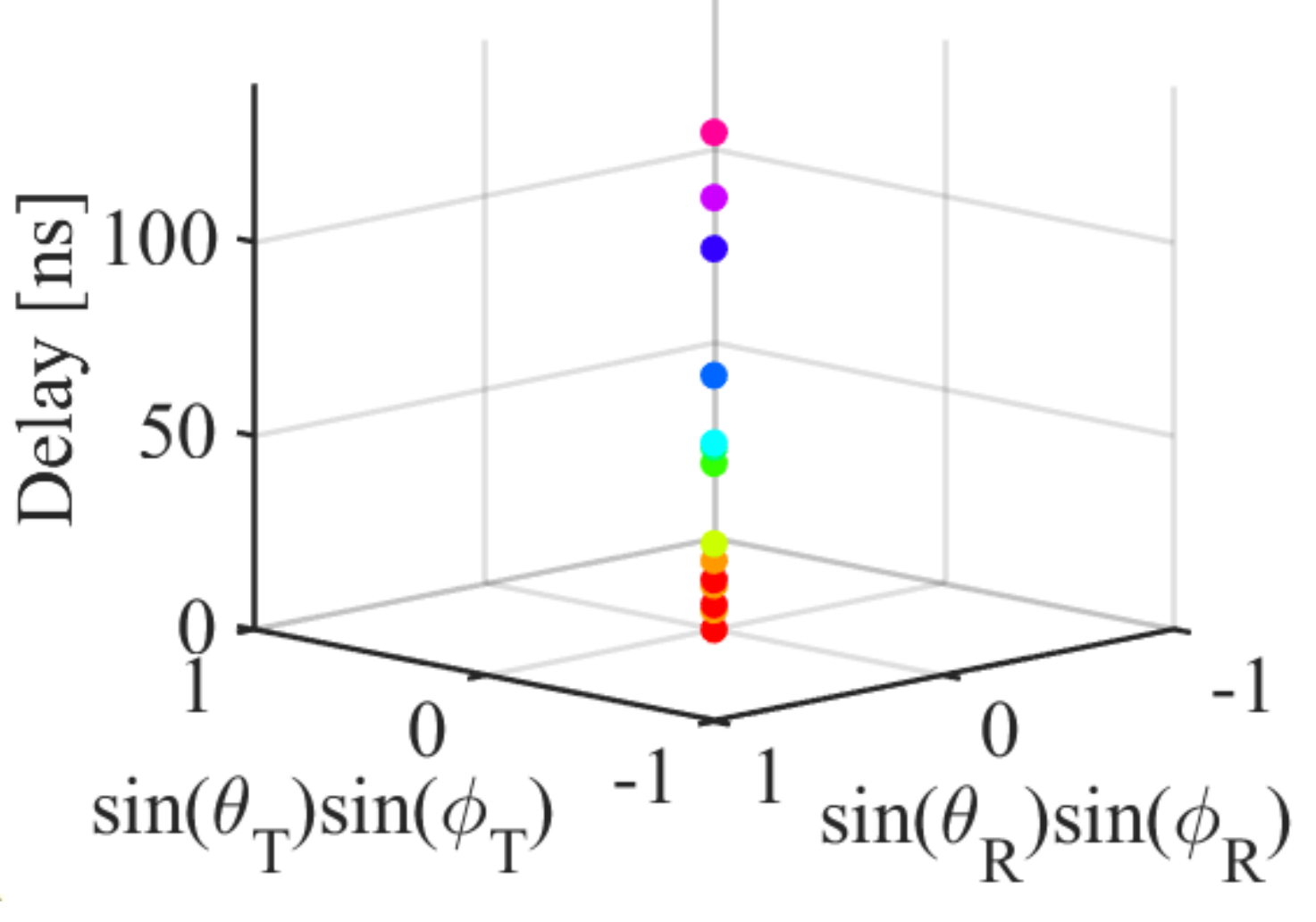}
\label{fig:old_model_1a_transformed} 
}
\subfloat[Transformed MPCs, $\hat{\mathbf{x}}_l$.]{
\includegraphics[width=0.23\textwidth]{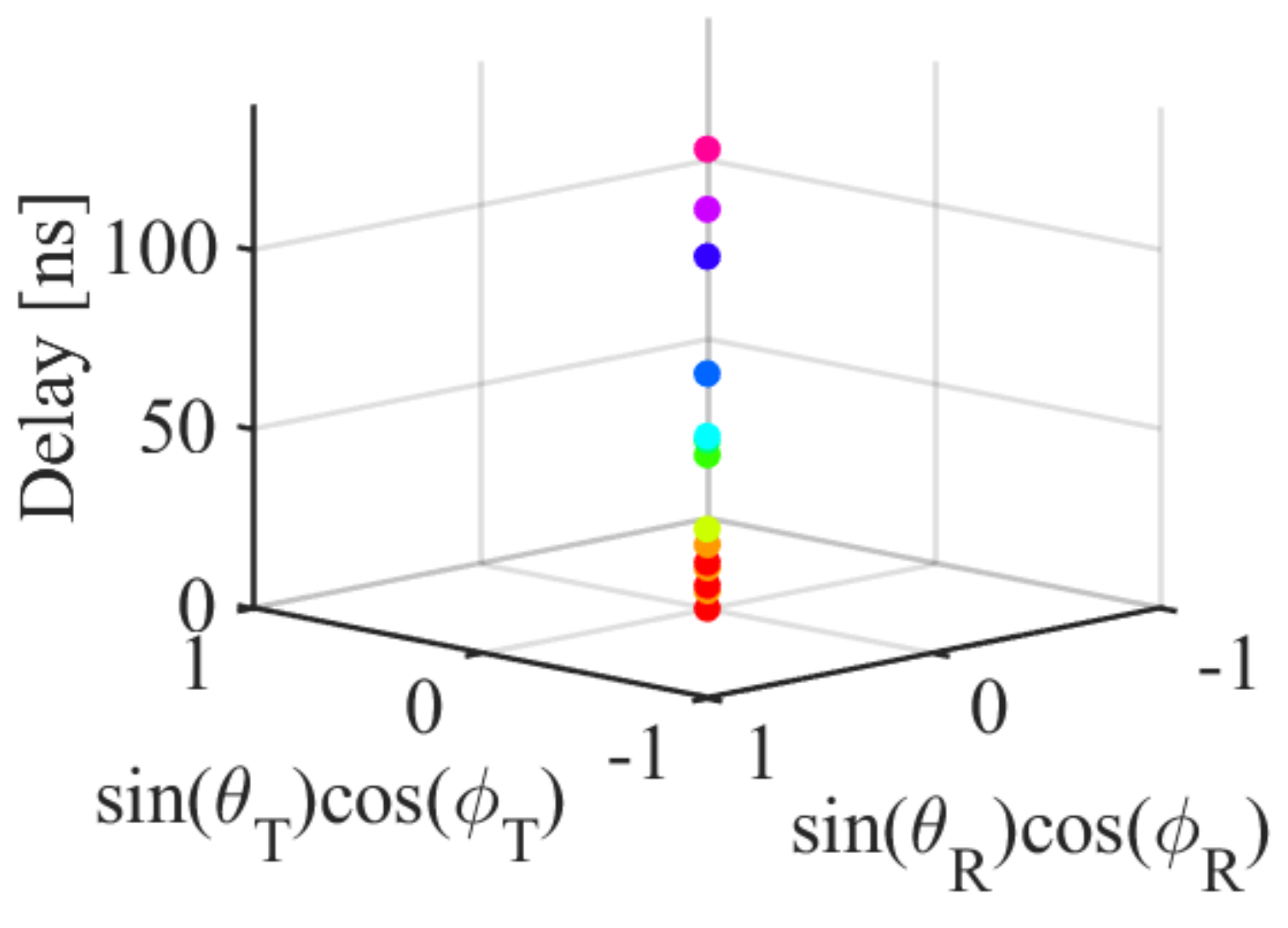}
\label{fig:old_model_1b_transformed} 
}
\caption{Original generated MPCs by 3GPP TR 38.901 model, $\mathbf{x}_l$, and the transformed MPCs via learned matrix $\mathbf{A}$, $\hat{\mathbf{x}}_l=\mathbf{A}^{\frac{1}{2}}\mathbf{x}_l$. 10 clusters are simulated. (a) and (b) Original generated MPCs. (c) and (d) Transformed MPCs after LMNN. }
\label{fig:old_model_mpcs}
\end{figure*} 
\par After learning $\mathbf{A}$ with MPCs generated the original 3GPP SCM, we observe that only the first element of the learned matrix $\mathbf{A}$ is non-zero, which is equivalent to an infinitely large value of delay weighting factor in MCD. Therefore, the learned distance metric by LMNN only considers the delay component of MPC. Recall that the proposed distance metric can be viewed as the Euclidean distance of the transformed MPCs by $\mathbf{A}^{\frac{1}{2}}$. In Fig.~\ref{fig:old_model_mpcs}, we carefully examine the original generated MPCs ( Fig.~\ref{fig:old_model_mpcs}(a) and (b)) and the transformed MPCs (Fig.~\ref{fig:old_model_mpcs}(c) and (d)), which demonstrate that the transformed MPCs are all located on the delay axis, and thereby validates that our previous analysis that only taking into account delay in distance metric would be preferred to improve clustering quality.
\subsection{A Modified Channel Model Based on 3GPP SCM}
\par To address the inappropriate modeling of intra-cluster power and intra-cluster delay in the current 3GPP channel model in evaluating clustering quality, we propose a modified version of the 3GPP channel model, as follows.
\begin{enumerate}
    \item The TOAs of intra-cluster MPCs are calculated as
    \begin{equation}
        \tau_{n,m+1}=\tau_{n,m}+\frac{1}{B_w},
    \end{equation}
    where $1/B_w$ represents the temporal resolution of the channel. This leads that the intra-cluster MPCs are resolvable in the delay domain.
    \item The power of the intra-cluster MPCs is exponentially decreased with intra-cluster TOA, AAOA and ZAOA, calculated as
    \begin{equation}
    \begin{split}
        p_{n,m}&=|\alpha_{n,m}|^2\\
        &=p_{n,1}\cdot\exp(-\gamma_\tau\frac{|\tau_{n,m}-\tau_{m,1}|}{|\tau_{n,M_n}-\tau_{m,1}|}\\
        &-\gamma_\theta\frac{|\theta_{n,m}-\theta_{n,1}|}{|\theta_{n,M_n}-\theta{m,1}|}\\
        &-\gamma_\phi\frac{|\phi_{n,m}-\phi_{n,1}|}{|\phi_{n,M_n}-\phi_{m,1}|}),
    \end{split}
    \end{equation}
    where $\gamma_\tau$, $\gamma_\theta$ and $\gamma_\phi$ denote the descent rates and are equal to $\frac{1}{3}\ln10$ in our simulation so that the weakest power of an MPC is one-tenth of the strongest one in a cluster.
    \item The power of intra-cluster MPCs is normalized as
    \begin{equation}
        p_{n,m}^\prime=\frac{p_{n,m}}{\sum_m{p_{n,m}}}P_n,
    \end{equation}
    where $P_n$ is the power of $n^{\text{th}}$ cluster.
    \item The TOAs of the intra-cluster MPCs are adjusted to satisfy the cluster delay spread, as
    \begin{equation}
        \tau_{n,m}^\prime=\frac{C_{\text{DS}}}{C^\prime_{\text{DS},n}}\tau_{n,m}+\tau_{\text{offset},n},
    \end{equation}
    where $C_{\text{DS}}$ stands for the cluster delay spread given in the 3GPP model and $C^\prime_{\text{DS}}$ states the calculated cluster delay spread of the $n^{\text{th}}$ cluster. In addition, $\tau_{\text{offset},n}$ is an offset that ensures the cluster delay, TOA of the first path in $n^{\text{th}}$ cluster, is unchanged after the adjustment, given by
    \begin{equation}
        \tau_{\text{offset},n}=\frac{C_{\text{DS}}^\prime-C_{\text{DS}}}{C^\prime_{\text{DS},n}}\tau_{n,1}.
    \end{equation}
\end{enumerate}
\begin{figure*}[htbp]
\subfloat[Original MPCs, $\mathbf{x}_l$.]{
\includegraphics[width=0.23\textwidth]{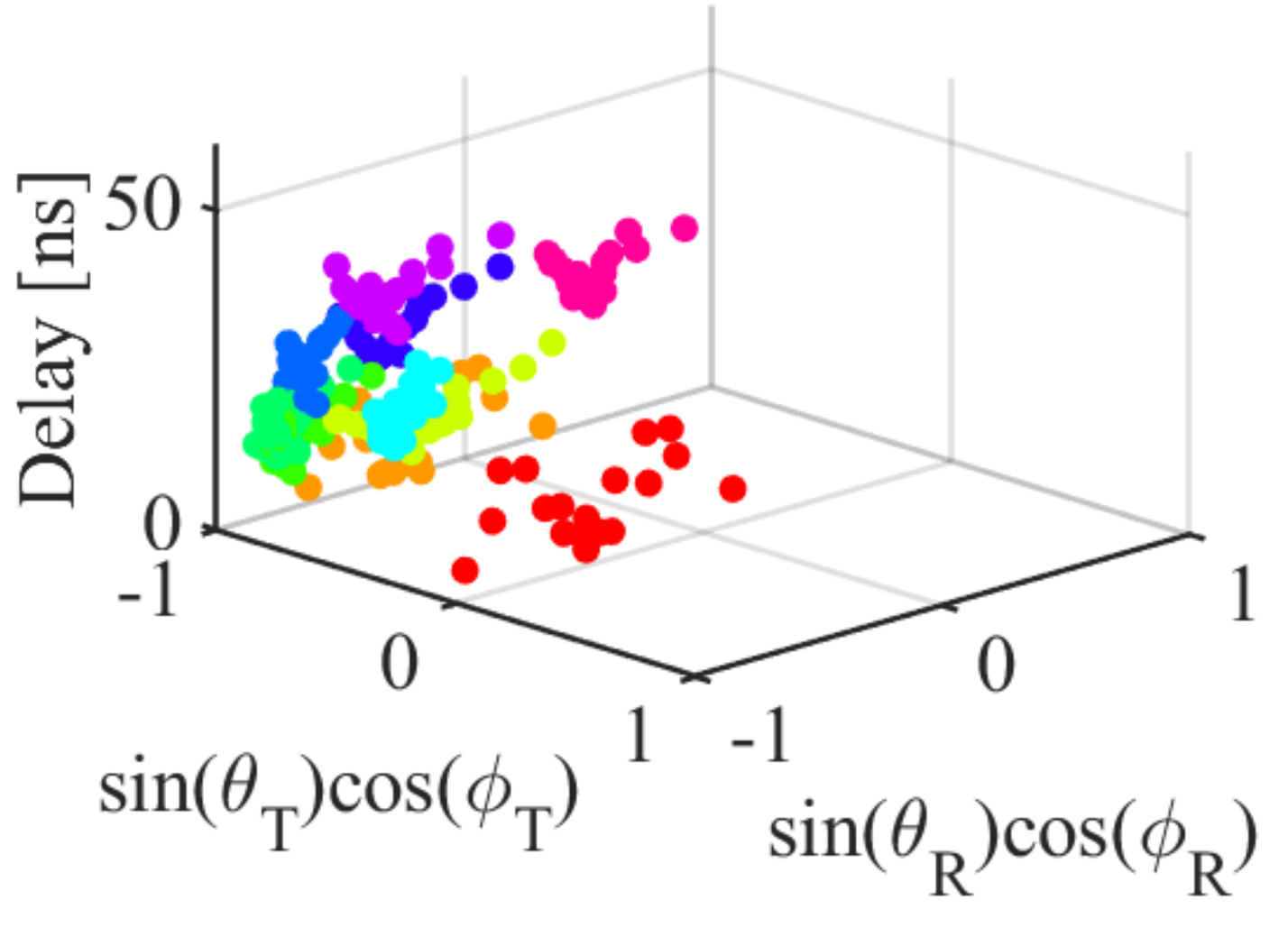}
\label{fig:new_model_1a}
}
\subfloat[Original MPCs, $\mathbf{x}_l$.]{
\includegraphics[width=0.23\textwidth]{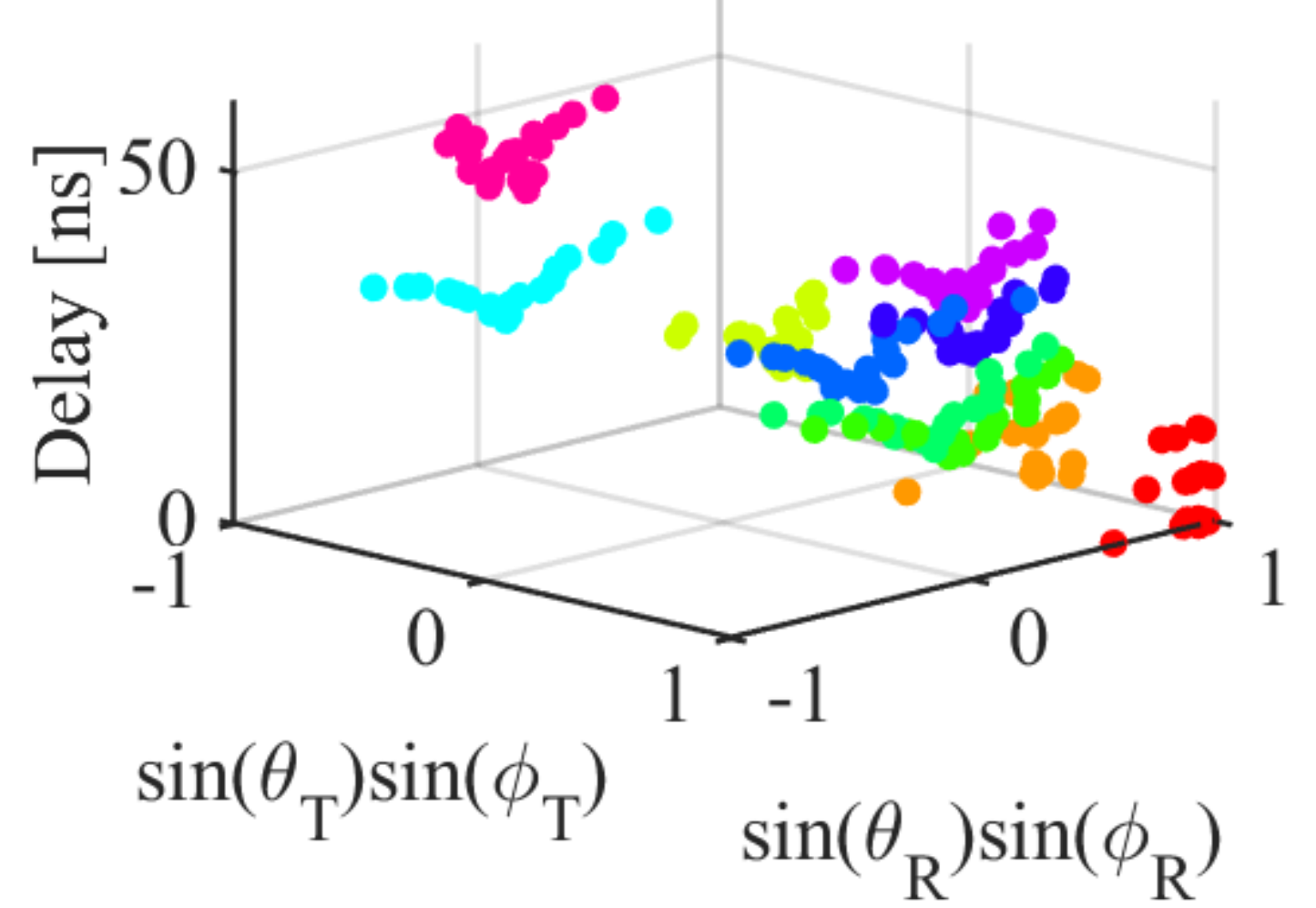}
\label{fig:new_model_1b} 
}
\subfloat[Transformed MPCs, $\hat{\mathbf{x}}_l$.]{
\includegraphics[width=0.23\textwidth]{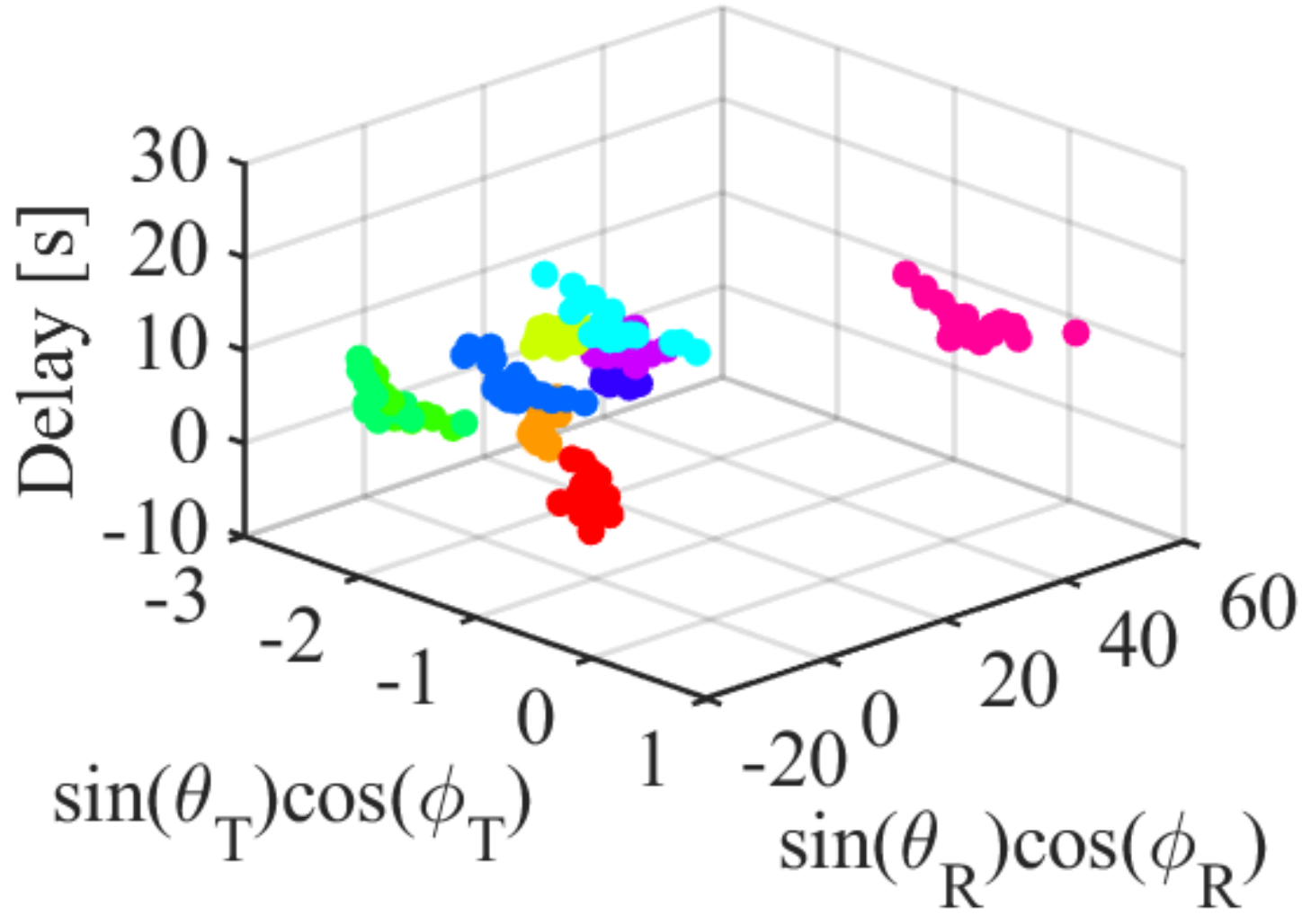}
\label{fig:new_model_1a_transformed} 
}
\subfloat[Transformed MPCs, $\hat{\mathbf{x}}_l$.]{
\includegraphics[width=0.23\textwidth]{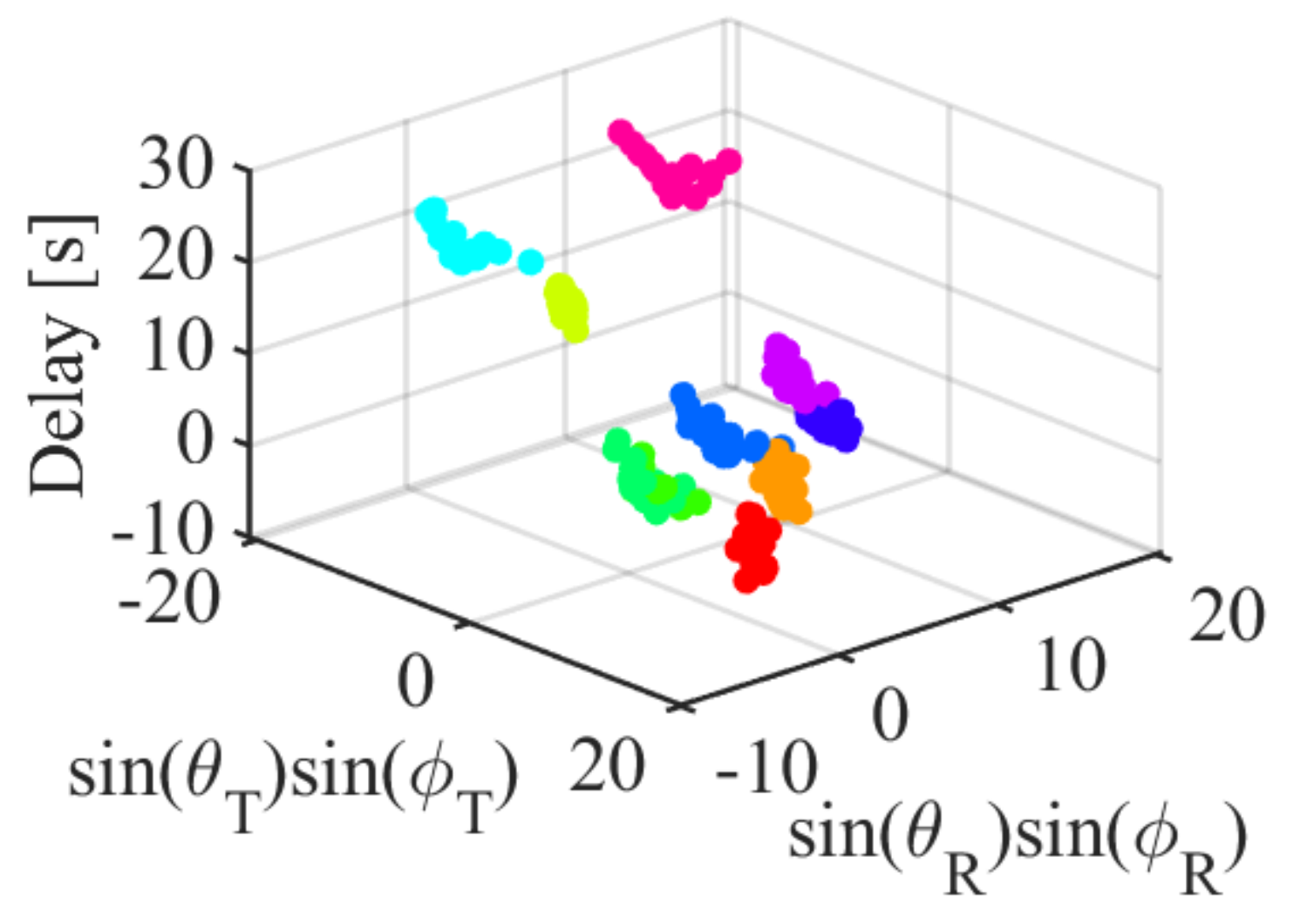}
\label{fig:new_model_1b_transformed} 
}
\caption{Original generated MPCs by our modified 3GPP model, $\mathbf{x}_l$, and the transformed MPCs via learned matrix $\mathbf{A}$, $\hat{\mathbf{x}}_l=\mathbf{A}^{\frac{1}{2}}\mathbf{x}_l$. 10 clusters are simulated. (a) and (b) Original generated MPCs. (c) and (d) Transformed MPCs after LMNN.}
\label{fig:new_model_mpcs}
\end{figure*} 

\par Figure~\ref{fig:new_model_mpcs} illustrates the generated MPCs by our modified 3GPP model and the transformed MPCs via learned matrix $\mathbf{A}$ by the LMNN algorithm. Compared with the transformed MPCs generated by 3GPP channel model in Fig.~\ref{fig:old_model_mpcs}(c) and~\ref{fig:old_model_mpcs}(d) that are located on the delay axis, the transformed MPCs generated by our modified models well spread in the delay-angle domain, which suggests that both delay and angle information are considered in the learned distance metric. \rev{More importantly, the transformed intra-cluster MPCs are more compact in Fig.~\ref{fig:new_model_mpcs}(c) and~\ref{fig:new_model_mpcs}(d) than the original MPCs in Fig.~\ref{fig:new_model_mpcs}(a) and~\ref{fig:new_model_mpcs}(b). Moreover, different clusters in Fig.~\ref{fig:new_model_mpcs}(c) and~\ref{fig:new_model_mpcs}(d) are more separate. }

\section{Simulation and Performance Analysis}
In this section, we present a modified reference channel model based on 3GPP SCM to generate the clustered MPCs. Next, the proposed Mahalanobis-distance metric learned by MMC and LMNN is analyzed, in comparison to the clustering quality of KMeans, KPowerMeans, and DBSCAN by using MCD.

\subsection{Simulation Setup}
We consider the MPCs with ground-truth labels generated from channel models for evaluating and comparing the clustering quality. In each experiment, we generate 200 channel realizations via the 3GPP TR 38.901 channel model~\cite{3gpp2018study} or the modified 3GPP-like channel model to calculate the averaged validation measure of the clustering results of three clustering algorithms, i.e., KMeans, KPowerMeans, and DBSCAN, based on MCD and learned distance metric with $\mathbf{A}$. In the experiment, the delay weighting factor in MCD is set to be 1, which is a reasonable value though it may not be optimal. In the channel realization, the carrier frequency $f_c$ is set to 60~GHz with the bandwidth $B_w=2$~GHz and the scenario is Urban Micro with line-of-sight (LoS), which is the same as in~\cite{he2017kernel}. The number of clusters, $N$, varies from 10 to 30 for comparison. 
In the learning process, we randomly pick $M_{\text{train}}$ MPCs of $N_{\text{train}}$ clusters from $S_{\text{train}}$ channel realizations to establish a training set. That is, $M_{\text{train}}N_{\text{train}}S_{\text{train}}$ MPCs with labels are trained to obtain the matrix $\mathbf{A}$. Unless specified, $N_{\text{train}}$ and $M_{\text{train}}$ are both equal to 5, whereas $S_{\text{train}}$ is 1. In the simulation figures, we use `MMC' to denote the clustering results from the clustering algorithms that use the distance metric learned by the MMC algorithm (i.e., learning the diagonal matrix $\mathbf{A}$), while `LMNN' to denote the use of the LMNN method (i.e., learning the full matrix $\mathbf{A}$).
\par In addition, the cases with and without AOD are separately considered in the simulation. In the case with AOD, the AAOD and ZAOD are involved in calculating the distance metric of MPCs. On the contrary, in the case without AOD, AAOD and ZAOD of MPCs are regarded to be unknown, which appears usually in the channel measurement at mmWave and higher frequency bands, e.g., the THz band~\cite{yu2020wideband}. This is due to the fact that scanning the antennas or shifting antenna to visualize antenna array at both Tx and Rx sides is extraordinary time-consuming. In addition, clustering MPCs with partial angle information of MPCs becomes more challenging than that with full angle information of MPCs.
\par For performance evaluation, the F measure, which is a robust external quality measure, is chosen as the merit of the goodness of the clustering results~\cite{amigo2009comparison}. We denote $C$ as the set of clusters from the clustering algorithms and $L$ as the set of categories of the measured clusters. The precision of a cluster $C_i$ is defied as
\begin{equation}
    \text{Precision}(C_i,L_j)=\frac{|C_i \cap L_j|}{|C_i|},
    \label{eq:precision}
\end{equation}
where $|C_i \cap L_j|$ is the number of paths that belong to $j^\mathrm{th}$ cluster and are identified into $i^\mathrm{th}$ cluster by clustering algorithm. The recall of cluster $L_i$ is given as
\begin{equation}
    \text{Recall}(L_i,C_j)=\text{Precision}(C_j,L_i).
    \label{eq:recall}
\end{equation}
\par Then F measure of cluster $L_i$ and $C_j$ is a harmonic average of \eqref{eq:precision} and \eqref{eq:recall} as
\begin{equation}
    F(L_i,C_j)=\frac{2\times \text{Recall}(L_i,C_j)\times \text{Precision}(L_i,C_j)}{ \text{Recall}(L_i,C_j)+ \text{Precision}(L_i,C_j)}
\end{equation}
\par Finally, the overall F measure is calculated as,
\begin{equation}
    F=\sum_{i=1}^N\frac{|L_i|}{N}\max_j\{F(L_i,C_j)\}.
\end{equation}
\par The F measure ranges from 0 to 1, where 1 indicates that the MPC clustering results are exactly the same as the ground-truth labels of the MPCs, i.e., perfectly clustering.
\subsection{Impact of Channel Models}
\begin{figure}
\centering
\includegraphics[width=0.4\textwidth]{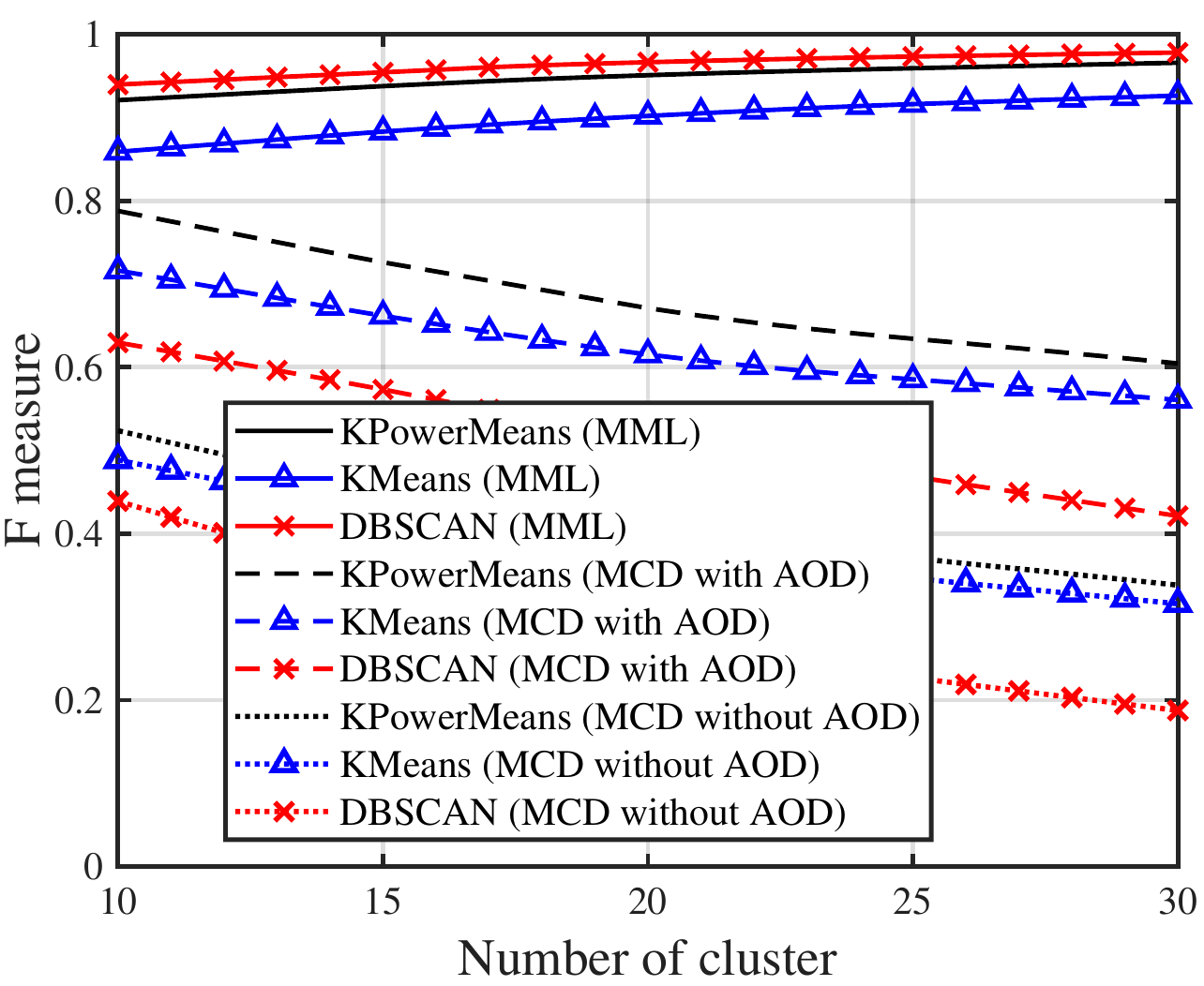}
\caption{F measure of clustering algorithms with original 3GPP TR 38.901 channel model by varying number of clusters, $N$. `MML' denotes the usage of distance metric obtained by Mahalanobis-distance metric learning wile MCD denotes taking MCD as distance metric. }
\label{fig:old_model_fmeasure}
\end{figure}

\begin{figure}[htbp]
\centering
\includegraphics[width=0.4\textwidth]{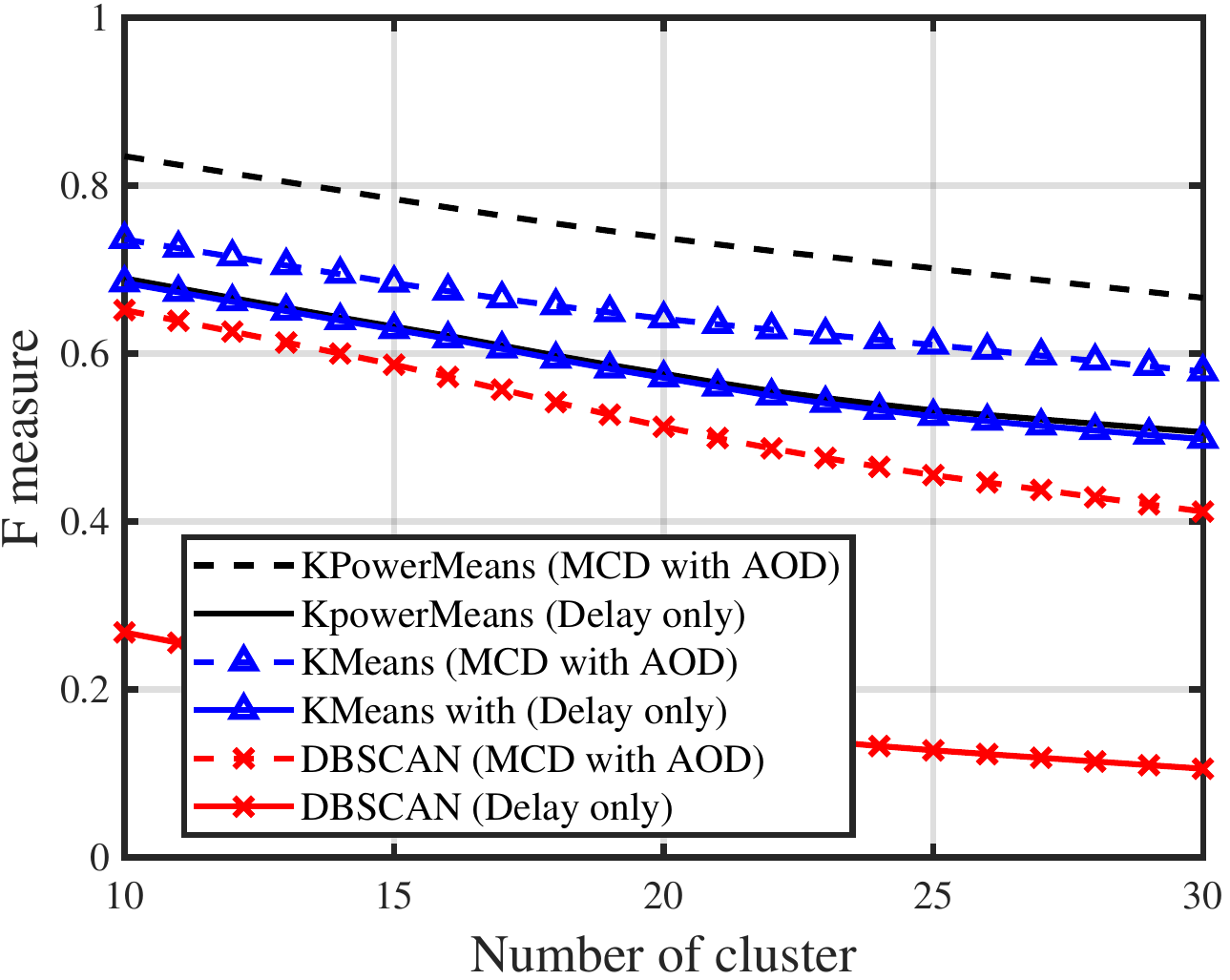}
\caption{F measure of clustering algorithms with our modified 3GPP channel model by varying number of clusters.
}
\label{fig:new_model_delayonly}
\end{figure}
\par With MPCs generated by 3GPP TR 38.901 channel model, Figure~\ref{fig:old_model_fmeasure} shows the F measure of different clustering algorithms using distance metric by metric learning and MCD as distance metric. Clustering algorithms using learned distance metric by metric learning that only considers delay component of MPCs shows significant improvement of clustering quality compared with those that use MCD as the distance metric. As the number of clusters increases from 10 to 30, the F measure with learned distance metric by LMNN increases, while the F measure with MCD decreases. To be concrete, KPowerMeans and DBSCAN using the learned distance metric have the F measure approximate to 1, when the number of clusters is 30. \rev{The reason for the increasing F measure is that the incorrectly-clustered MPCs only occur in the two clusters with the strongest cluster power (we have discussed in Sec.~IV-A) and the ratio of perfectly-clustered MPCs among all the MPCs tends higher as the total number of clusters increases. For example, $80\%$ MPCs are perfectly clustered with $N=10$ and $90\%$ MPCs are perfectly clustered with $N=20$.} On the contrary, the reason for the decreasing F measure of MCD is that when more clusters come into space, the clustering quality becomes worsen if angle information is taken into account. Furthermore, we observe that the F measure of MCD without AOD is much lower than that of MCD with AOD. It should be noted that since only the delay is considered, F measures of LMNN with AOD and without AOD have no difference.
\par To study our modifications on the 3GPP channel model, we calculate the F measure of different clustering algorithms with MPCs generated by our modified channel model, by varying the number of clusters, $N$, as demonstrated in Fig.~\ref{fig:new_model_delayonly}. `Delay only' denotes that the distance metric only takes the distance component of MPCs into account, which shows near-optimal clustering quality on the original 3GPP channel model in Fig.~\ref{fig:old_model_fmeasure}, while `MCD with AOD' denotes taking MCD with AOD information as the distance metric. It can be observed that if the delay component of MPCs is only considered in the distance metric, the F measure would be lower than that by taking MCD with AOD as the distance metric. Also, the F measure decreases with the increasing $N$. These suggest that our modified channel model has addressed the inappropriate features of the original 3GPP channel model in MPC clustering quality evaluation.

\begin{figure}
\subfloat[Impact of snapshot number for training, $S_{train}$, given $N_{train}$ and $M_{train}$ are 5.]{
\includegraphics[width=0.4\textwidth]{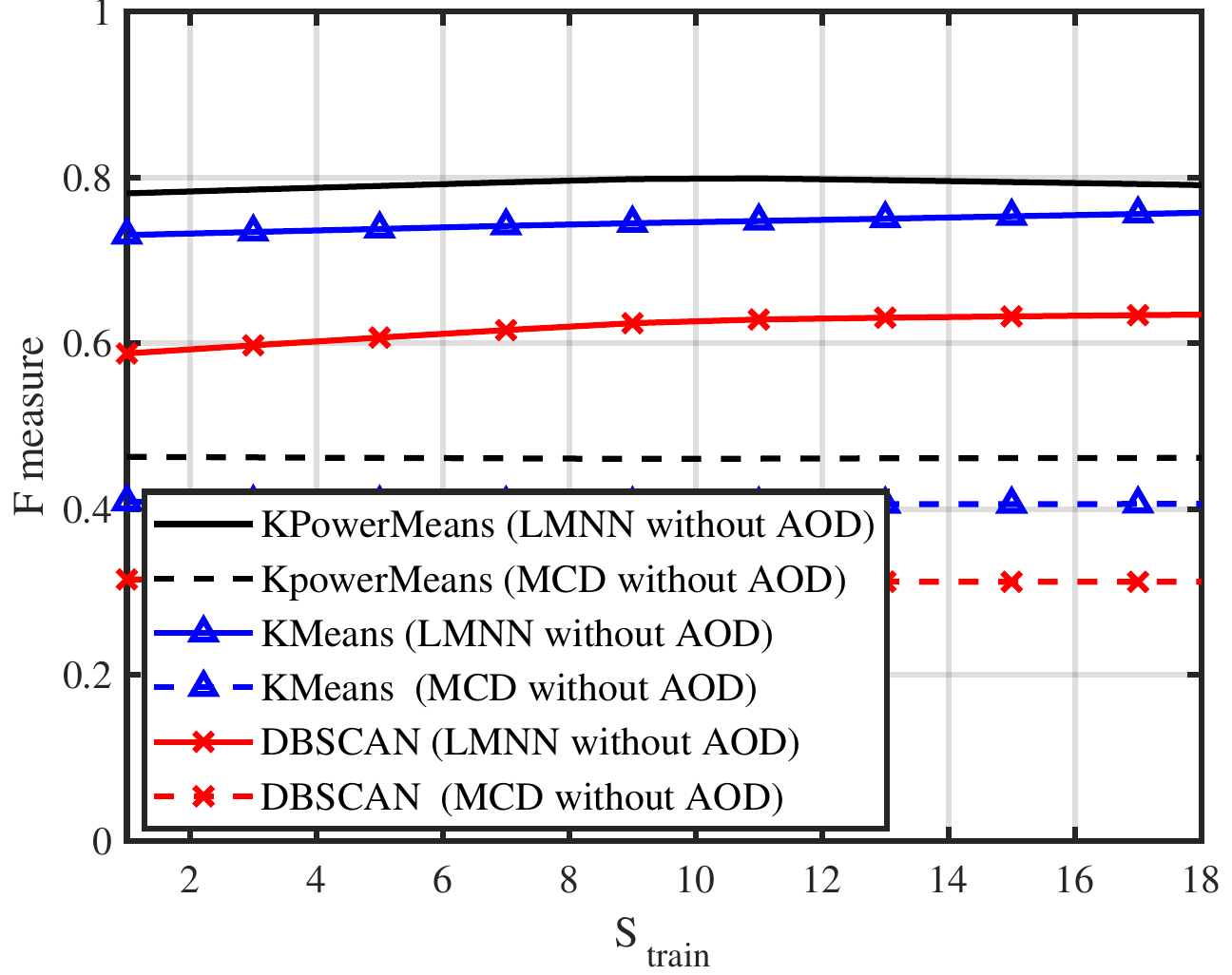}
\label{fig:strain}
}

\subfloat[Impact of intra-cluster path number for training, $M_{train}$, given $N_{train}$ is 5 and $S_{train}=1$.]{
\includegraphics[width=0.42\textwidth]{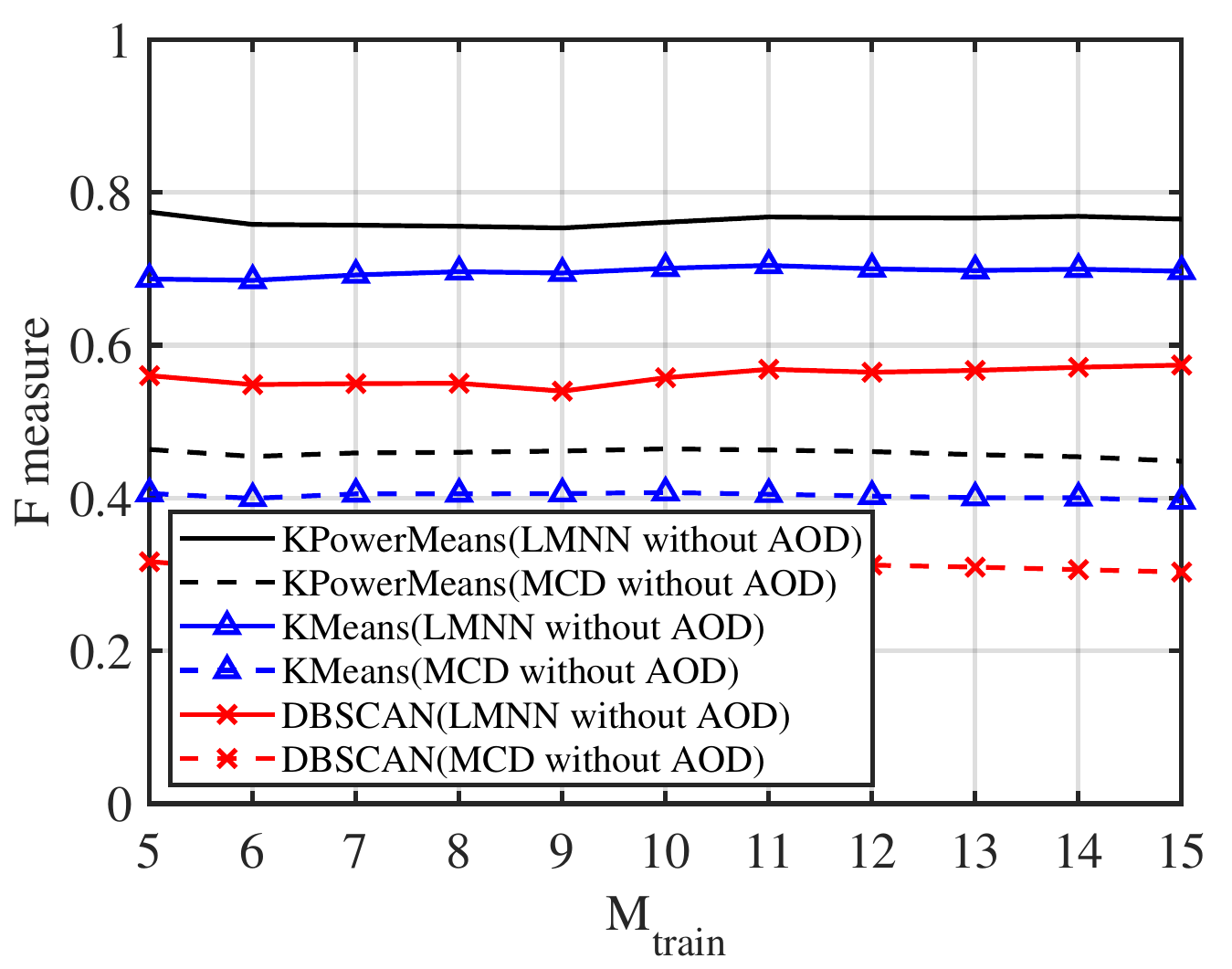}
\label{fig:mtrain} 
}
 
\subfloat[Impact of cluster number for training, $N_{train}$, given $M_{train}$ is 5 and $S_{train}=1$.]{
\includegraphics[width=0.4\textwidth]{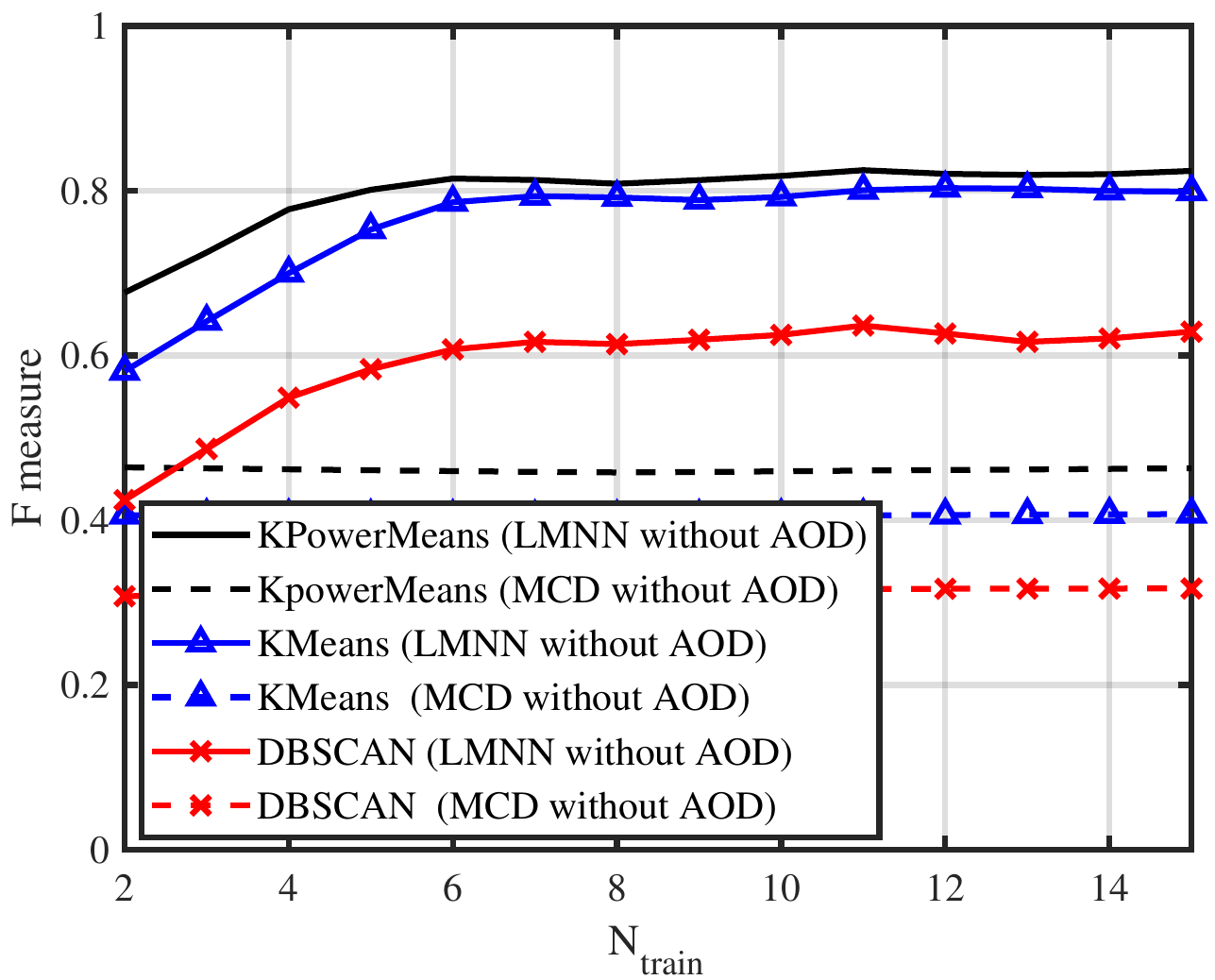}
\label{fig:ntrain} 
}

\caption{Impact of snapshot number, $S_{\text{train}}$, intra-cluster path number, $M_{\text{train}}$ and cluster number, $N_{\text{train}}$ for training.}
\label{fig:learning_analysis}
\end{figure} 

\subsection{Learning Analysis}
\rev{To understand how many labeled MPCs are needed for metric learning, we calculate the F measure by varying the number of snapshots for training, $S_{train}$, and the number of clusters for training, $N_{train}$, respectively. In the simulation, 20 clusters in each snapshot are simulated.
Moreover, the LMNN algorithm is used to learn the full $\mathbf{A}$, while AOD is not considered in the distance metric. Figure~\ref{fig:learning_analysis}(a) evaluates the impact of $S_{train}$ with the fixed $N_{train}$ and $M_{train}$ of 5. The results show that the F measure is not sensitive to $S_{train}$, which suggests that only 25 MPCs from one snapshot are sufficient for metric learning.
Figure~\ref{fig:learning_analysis}(b) shows that the F measure is not sensitive to the number of intra-cluster paths for training, $M_{train}$, the when $M_{train}$ exceeds 5 given $S_{train}=1$ and $N_{train}=5$. In addition, Fig.~\ref{fig:learning_analysis}(c) presents the influence of $N_{train}$ with the fixed $S_{train}=1$ and $M_{train}=5$. We observe that the F measure of KPowerMeans tends to be stable after $N_{train}$ exceeds 5, while this saturating number of $N_{train}$ for KMeans and DBSCAN equals to 6. As a result, we suggest that 5 labeled intra-cluster MPCs from 5 clusters of one snapshot, i.e., totally 25 labeled MPCs, are required to have near-optimal performance for metric learning, with low training overhead. }
\subsection{Impact of Number of Clusters}
\begin{figure}[htbp]
\subfloat[Diagonal $\mathbf{A}$ learned by MMC.]{
\includegraphics[width=0.4\textwidth]{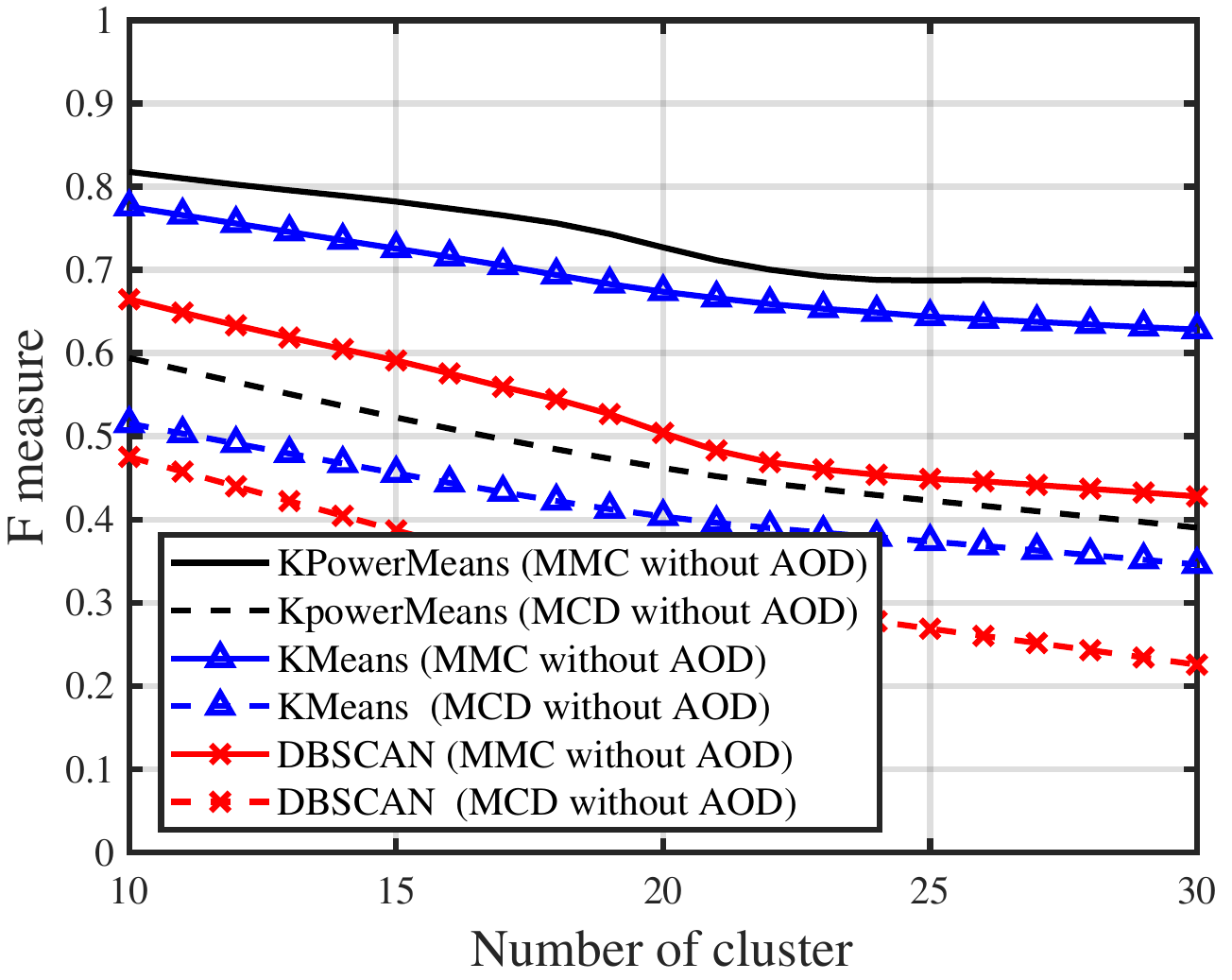}
\label{fig:new_fmeasure_mmc}
}

\subfloat[Full $\mathbf{A}$ learned by LMNN.]{
\includegraphics[width=0.4\textwidth]{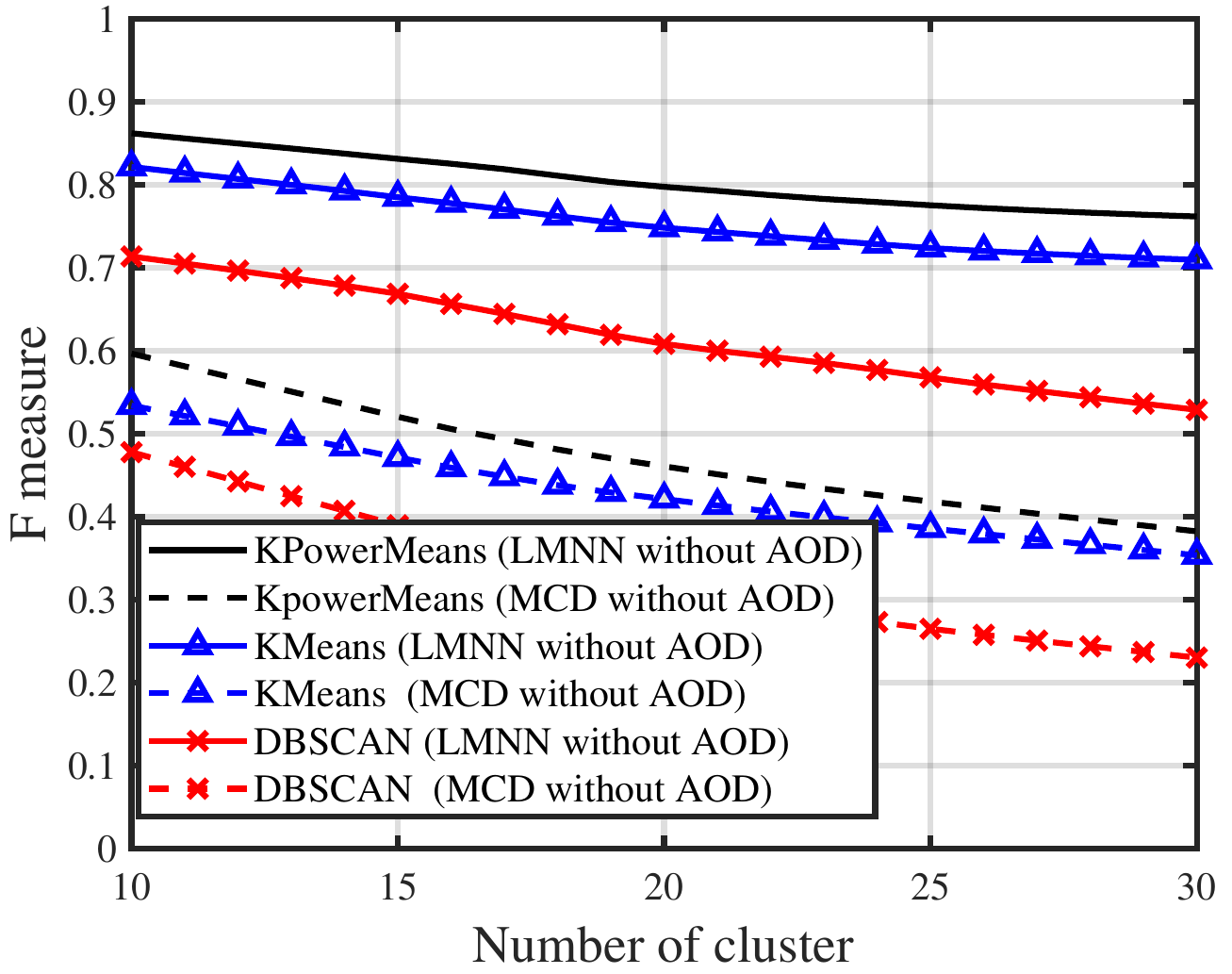}
\label{fig:new_fmeasure_lmnn} 
}
\caption{Impact of cluster. AOD is not considered in distance metric.}
\label{fig:new_fmeasure}
\end{figure}

The number of clusters, $N$, is a parameter that affects the clustering quality. Generally, the more cluster, the worsen the clustering quality. Figure~\ref{fig:new_fmeasure}(a) and (b) depict the F measure with the diagonal matrix $\mathbf{A}$ learned by MMC and full matrix $\mathbf{A}$ learned by LMNN respectively, when AOD is not considered in the distance metric. First, the diagonal matrix $\mathbf{A}$ can significantly improve F measures of KPowerMeans, KMeans, and DBSCAN, respectively. For example, when $N$ increases from 10 to 30, the F measure of KPowerMeans using MCD as distance metric decreases from 0.6 to 0.4, while that of KPowerMeans based on learned distance metric learned by MMC decreases from 0.82 to 0.68, which presents that an improvement of at least 0.2 for the F measure is achieved. Second, the full matrix $\mathbf{A}$ can further improve the clustering quality, compared with diagonal A. We take KPowerMeans as an example, in which the F measure by using learned full $\mathbf{A}$ is 0.86 and 0.76 when $N$ is 10 and 30, respectively.
\par
\begin{figure}[htbp]
\subfloat[Diagonal $\mathbf{A}$ learned by MMC.]{
\includegraphics[width=0.4\textwidth]{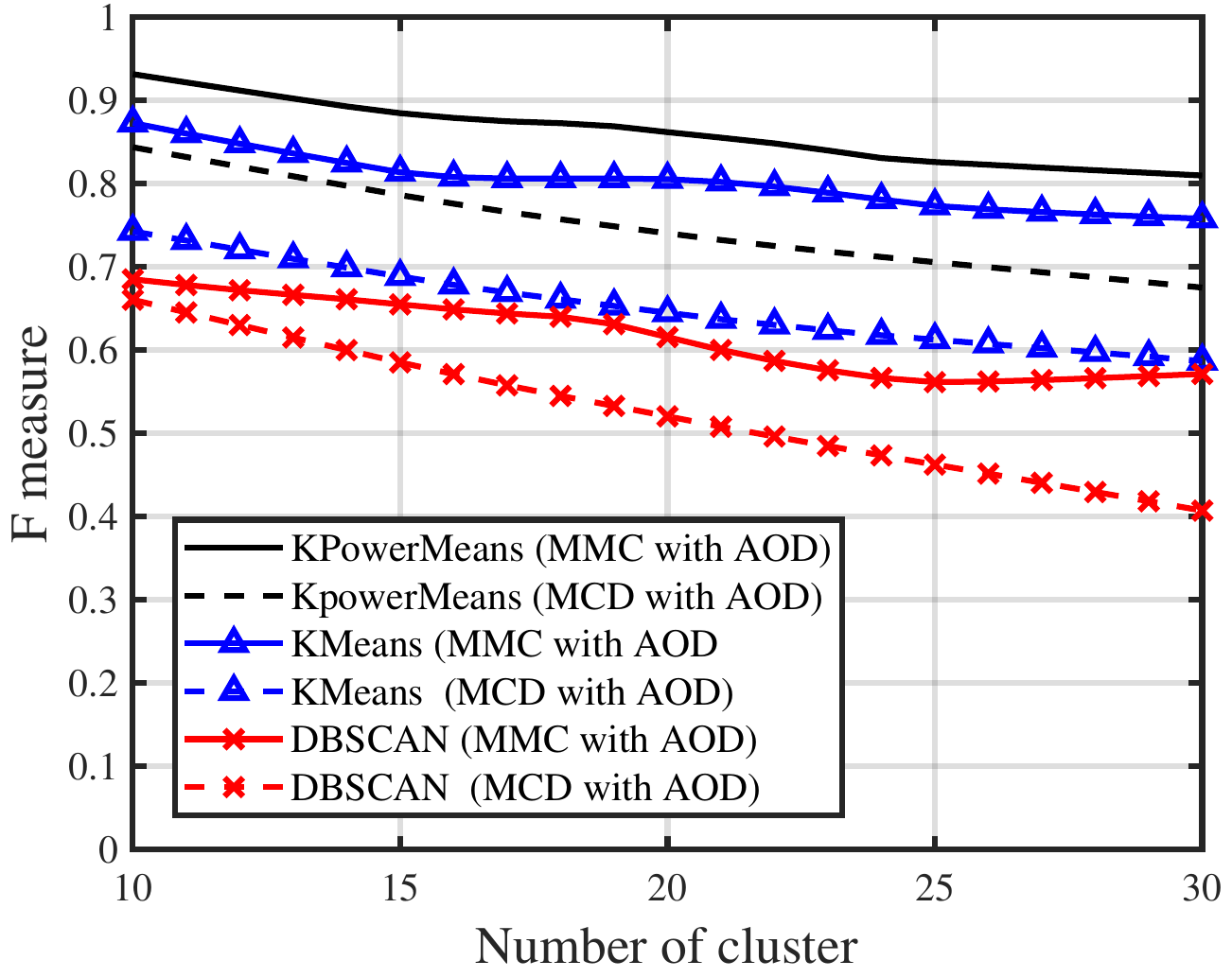}
\label{fig:new_fmeasure_mmc_aod}
}

\subfloat[Full $\mathbf{A}$ learned by LMNN.]{
\includegraphics[width=0.4\textwidth]{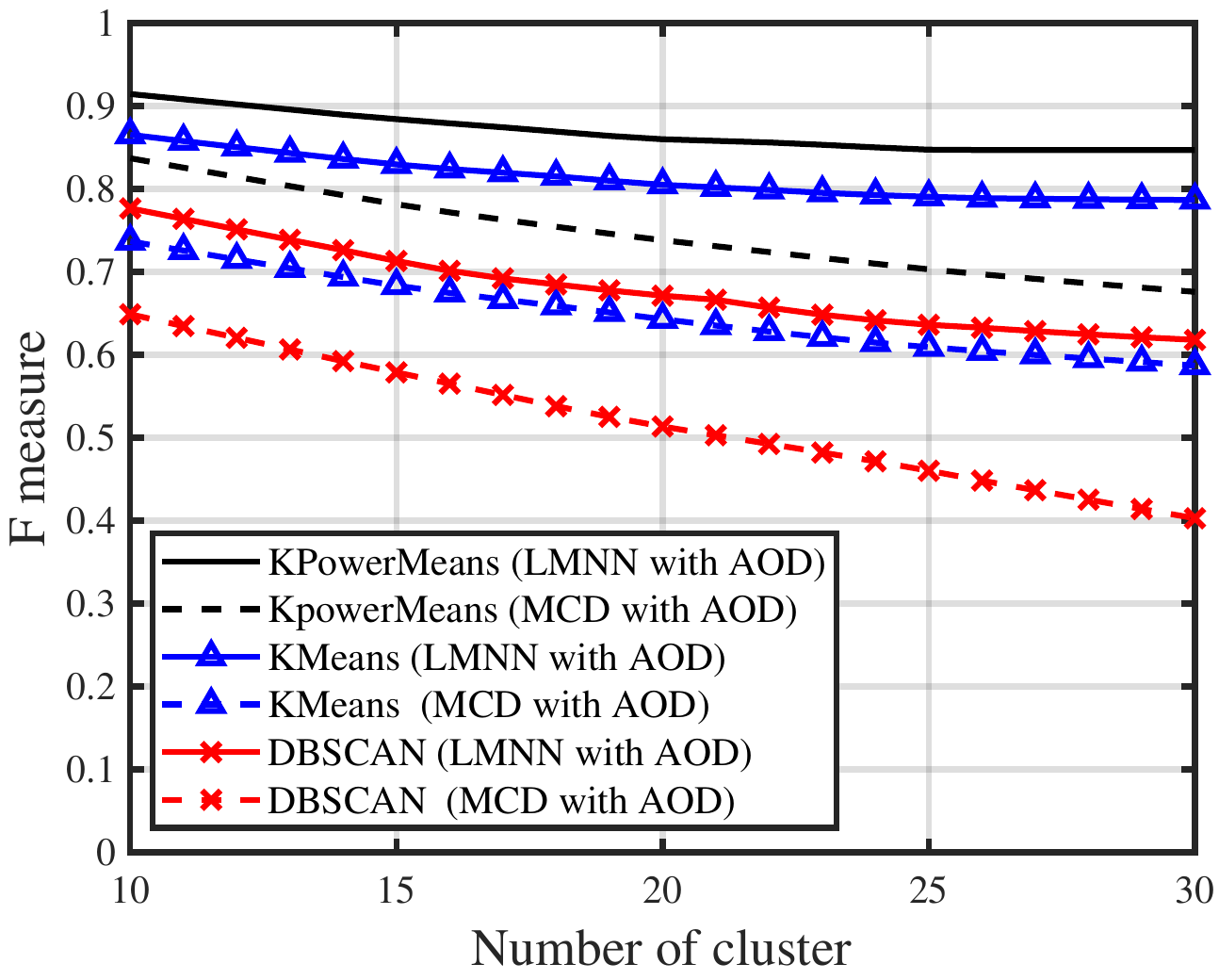}
\label{fig:new_fmeasure_lmnn_aod} 
}
\caption{Impact of cluster. AOD is considered in distance metric.} 
\label{fig:new_fmeasure_aod}
\end{figure}

 Similar observations are perceived in the case with AOD, as demonstrated in Fig.~\ref{fig:new_fmeasure_aod}. Especially, the improvement of clustering quality by full matrix $\mathbf{A}$ is more noticeable when the number of clusters is large. We observe that the improvement of F measure with learned distance metric due to the knowledge of AOD is not as significant as the improvement of F measure with MCD. This is due to the fact that MCD does not make full use of the delay and AOA information of MPCs in clustering, while the learned distance metric can learn the cluster features from the labeled MPCs and exploit delay and AOA information of MPCs. This is verified by an interesting observation that the F measure of KPowerMeans enabled by learned distance without considering AOD is even larger than that by using MCD that contains the AOD information.

\subsection{Impact of Angular Spared}
\begin{figure}
\centering
\includegraphics[width=0.4\textwidth]{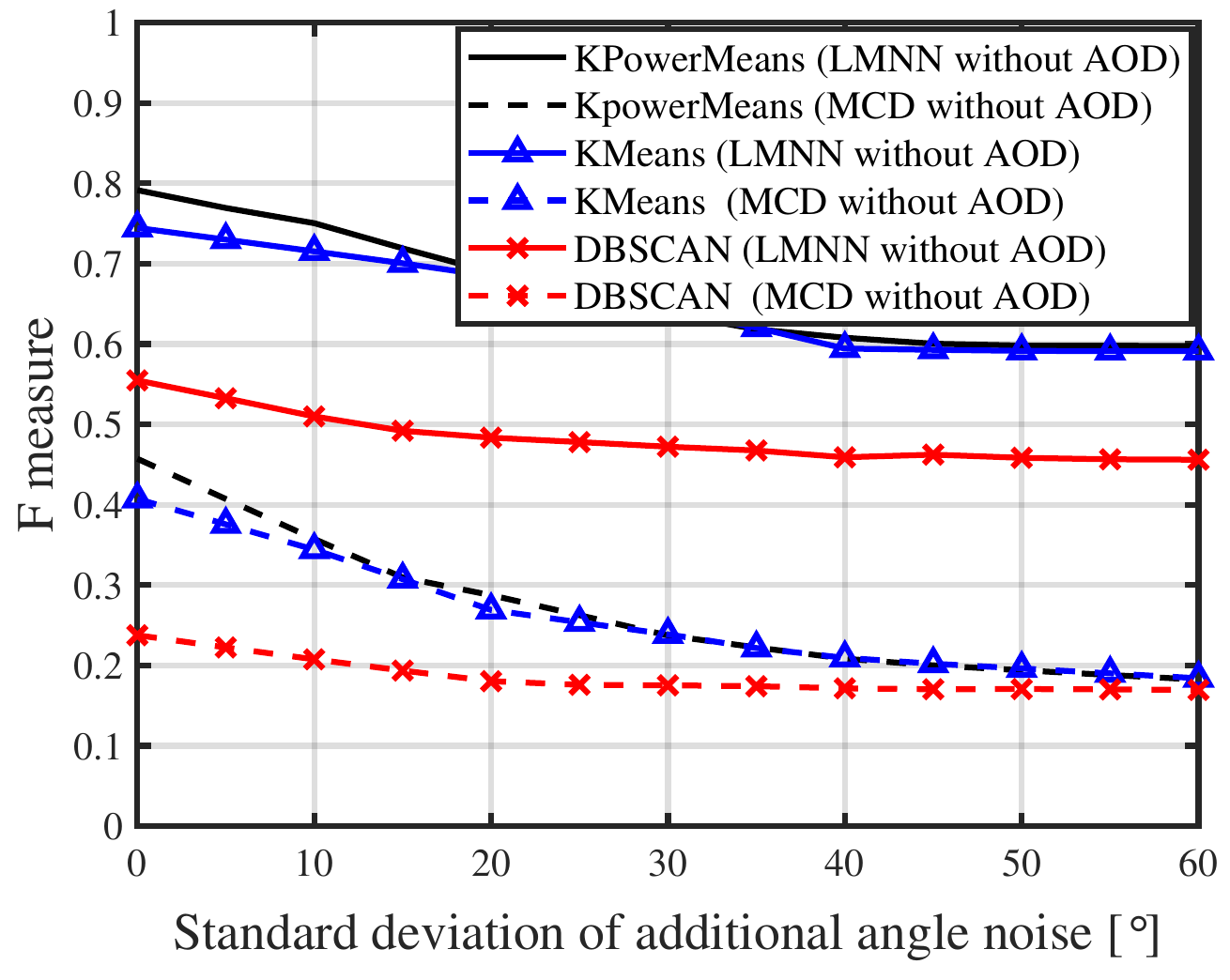}
\caption{Impact of angular spread.}
\label{fig:angle_noise_lmnn}
\end{figure}

To study the impact of angular spread on clustering quality, we add Gaussian white angle noise to the generated MPCs~\cite{czink2006framework,he2017kernel}. In the simulation, the standard deviation of the additional angle noise ranges from $0^\circ$ and $60^\circ$, and AOD is not considered in the distance metric. Figure~\ref{fig:angle_noise_lmnn} evaluates the F measure with varying angle noise. When the standard deviation of the angle noise reaches $40^\circ$, the F measure of three clustering algorithms that adopt MCD as distance metric all decrease to 0.2, which indicates very poor clustering quality. By contrast, the KPowerMeans and KMeans based on the LMNN-enabled learned distance metric maintain the F measure at a high level of 0.6. We further notice that the F measure using learned distance metric by LMNN with large angular spread is consistent with the F measure with `Delay only' illustrated in~\ref{fig:new_model_delayonly}, which suggests that the learned distance metric only takes delay into account. The reason is that the angle components of MPCs mislead the clustering algorithms when the angular spread is undesirably large. Therefore, the MCD suffers from the unnecessary consideration of angle components, while the distance metric by LMNN can learn this feature and intelligently adjust to the cluster characteristics of MPCs, leading to superior clustering quality.


\color{black}
\section{Validation with Channel Measurements}
In this section, we introduce the channel measurement campaign in a meeting room at 140 GHz and validate the proposed distance metric for MPC clustering with the measured MPCs.
\subsection{Channel Measurement Campaign}
\begin{figure}
\centering
\includegraphics[width=0.4\textwidth]{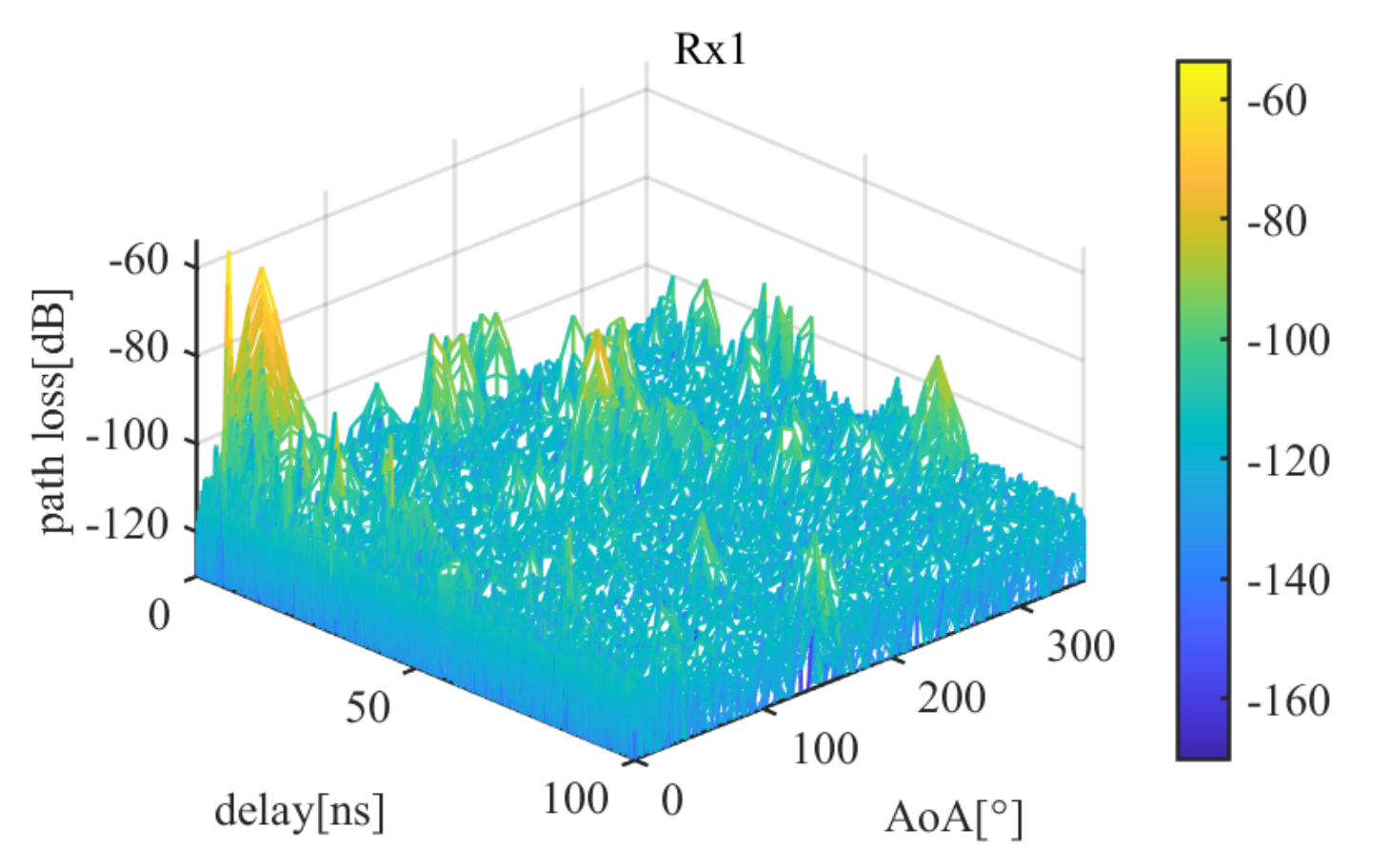}
\caption{\textcolor{black}{Measured power delay angular profile for Rx position 1.}}
\label{fig:pdap}
\end{figure}

A channel measurement campaign from 140~GHz to 143~GHz is conducted in a meeting room~\cite{chen2021channel}. The measured bandwidth is 13~GHz and the frequency interval is 10~MHz. In each measurement set, the position and orientation of the Tx antenna are fixed while the Rx position is moved. Rx antenna with a half-power beamwidth of $10^\circ$ is rotated in azimuth angle domain from $0^\circ$ to $360^\circ$ and elevation angle domain from $-20^\circ$ to $20^\circ$ with a step of $10^\circ$. In measurement set 1 of the channel measurement campaign, 10 Tx-Rx pairs are measured. Figure~\ref{fig:pdap} shows the power delay angular profile of Rx position 1. The threshold for filtering out the noise floor is 10~dB higher than the average noise power. For each measured channel, only the clusters with more than 15 multipaths are remained to form training and test datasets.

\subsection{Performance Validation with Measured MPCs}
\begin{figure}
\centering
\includegraphics[width=0.4\textwidth]{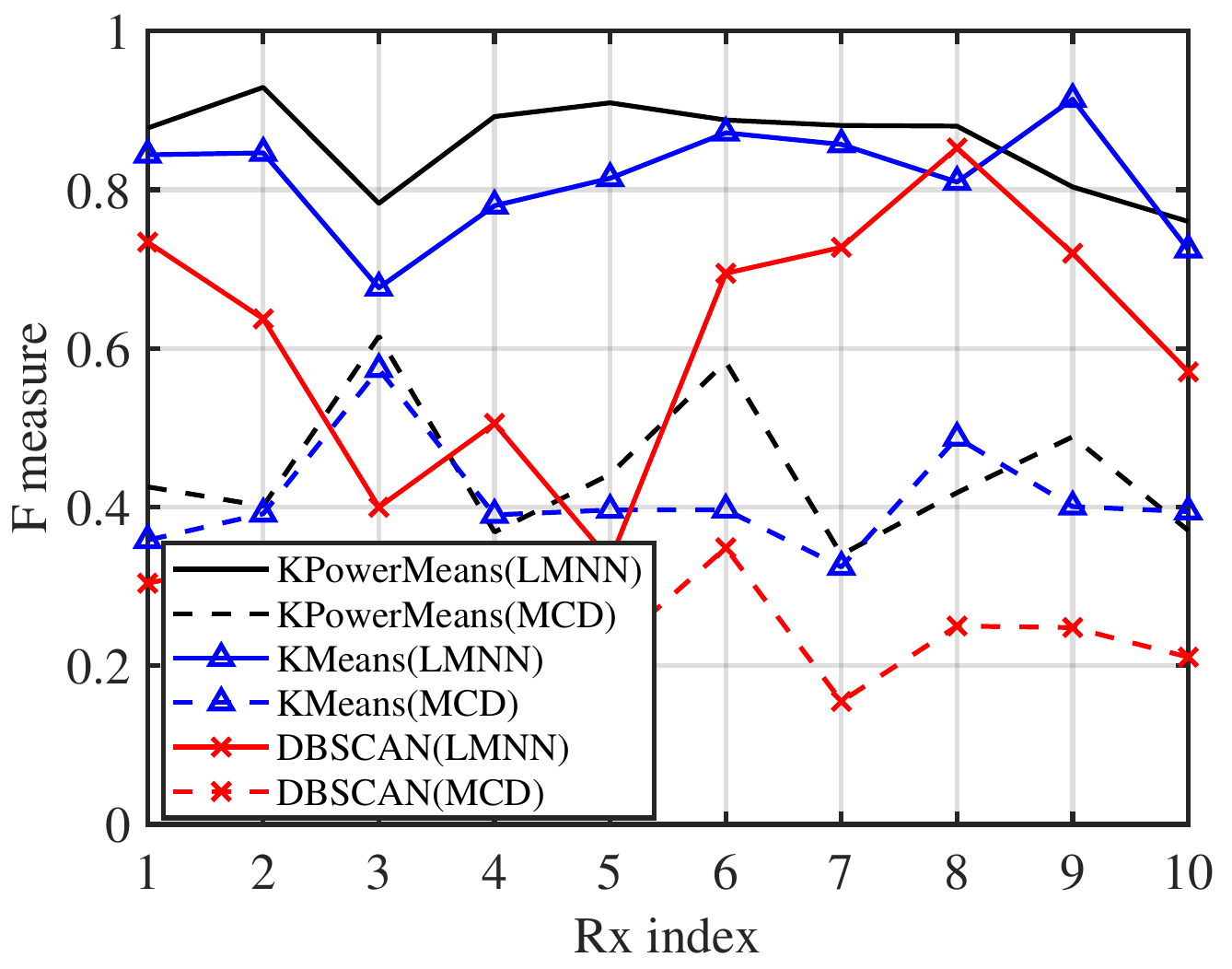}
\caption{\textcolor{black}{Validation with measured MPCs.}}
\label{fig:vali_measurement}
\end{figure}

To validate the proposed distance metric framework, we evaluate the F measure over the 10 measured channels with the learned distance metric (denoted as `LMNN') and MCD (denoted as `MCD') in Fig.~\ref{fig:vali_measurement}, respectively. The delay weighting factor in MCD is simply set as 1. The training set consisting of 25 MPCs randomly chosen from 5 clusters (5 MPCs from each cluster) in a Tx-Rx pair. The averaged F measure values for three clustering algorithms, i.e., KPowerMeans, KMeans, and DBSCAN, with the learned distance metric are 0.87, 0.83, and 0.67, respectively. By comparison, the F measure values with MCD are 0.43, 0.41, and 0.29, respectively, which suggests that the proposed distance metric can double the clustering quality in the real sub-THz channel measurement data.

\color{black}
\section{Conclusion and Open Problems}
\subsection{Conclusion}
In this paper, we propose a general framework of the Mahalanobis-distance metric for MPC clustering in MIMO channel analysis. Furthermore, we prove that the widely-used MCD in the literature is a special case of the proposed distance metric framework. The proposed distance metric for MPCs can be described by a semi-definite matrix $\mathbf{A}$. The usage of the proposed distance metric is equivalent to the Euclidean distance of the MPCs transformed via matrix $\mathbf{A}$, which presents good compatibility of the proposed distance metric with the existing algorithms. Two machine learning approaches with different designs of the loss function, i.e., MMC and LMNN, are developed to learn diagonal matrix$\mathbf{A}$ and full matrix $\mathbf{A}$, respectively. 
\par When evaluating MPC clustering quality, the original 3GPP SCM is found to be inappropriate due to the two following features. First, the power is equally distributed in a cluster. Second, MPCs in a cluster share the same TOA except for two clusters with the strongest power. Instead, we propose a modified channel model based on the original 3GPP SCM. The experiment results illustrate that the consideration of delay in the distance metric can achieve near-optimal clustering quality if MPCs are generated by the original 3GPP SCM model. By contrast, the proposed modified model appropriately addresses this problem. 

Based on MPCs generated by the modified channel model, we validate our proposed general framework of the Mahalanobis-distance metric. The evaluation results show that, in the training phase of the distance metric learning, 25 MPCs from 5 clusters in one snapshot are sufficient to achieve optimal clustering quality, which suggests very limited effort of manually labeling MPCs. More importantly, the learned distance metric outperforms the well-known MCD and largely improves the clustering quality of existing clustering algorithms. For example, if delay and AOA are known and the number of clusters is 30, the F measure of KPowerMeans by using the learned distance metric is 0.76, significantly better than the F measure of 0.38 by using MCD. It is worth noting that the F measure of KPowerMeans by using MCD is 0.68 with 30 clusters and the knowledge of AOD, which demonstrates that learned distance metric can compensate for the loss of clustering quality due to the partial knowledge of angle information of MPCs. This suggests that a good distance metric fully exploiting the information of MPCs is important in clustering MPCs.
The proposed distance metric framework is also validated by real sub-THz channel measurement data. The results show that the clustering quality with MCD is doubled by the learned distance metric compared.
\subsection{Open Problems}
As an attempt to introduce supervised learning to MPC clustering in MIMO channel analysis, the proposed distance metric for MPC is equivalent to the Euclidean distance of linear transformed MPCs. However, the \textit{non-linear transform of MPCs}, which is enabled by non-linear kernel methods and neural networks in machine learning, has the potential to further improve the clustering quality. The non-linear transform can be invoked in the proposed distance metric by introducing non-linear basis functions as described in Sec. III-A. However, the design of the basis functions relying on the cluster features requires many efforts including the analysis of well-clustered MPCs. 
\par An alternative approach to enable non-linear transform of MPCs is to use neural networks to replace the current distance metric in clustering algorithms. A possible method is \textit{Siamese Network} by combining two networks with the same structure, the input, and output of which are a couple of MPCs and the similarity of the input MPCs, respectively. A main challenge of the neural network method is the requirement of large-scale training datasets. The labeled training datasets are suggested to be obtained by visual inspection and thereby, very time-consuming. The reason why using clustering algorithms to obtain training datasets is not suggested is that the clustering results are determined by the inherent criteria of the clustering algorithms as well as the utilized distance metric. Therefore, the outcome of a trained neural network could be far from reality and its performance may be limited by the upper bound of the clustering algorithms. It is still under investigation how to fully exploit machine learning algorithms to further benefit clustering MPCs.

\bibliographystyle{IEEEtran}
\bibliography{IEEEabrv,CY_bib,bibliography.bib}
\vfill

\end{document}